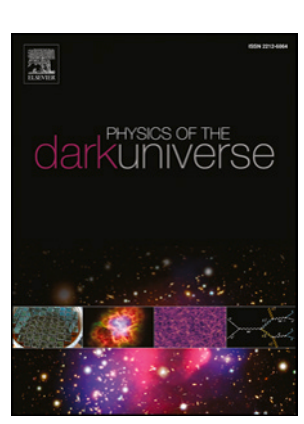

Full Length Article

# Constraints on axion-like particles from ultra-high-energy observations of M87 with the HAWC observatory

R. Alfaro [a,*], C. Alvarez [b], A. Andrés [c], E. Anita-Rangel [c], M. Araya [d], J.C. Arteaga-Velázquez [e], D. Avila Rojas [c,*], H.A. Ayala Solares [f], R. Babu [g], P. Bangale [f], E. Belmont-Moreno [a], A. Bernal [c], K.S. Caballero-Mora [b], T. Capistrán [c,h], A. Carramiñana [i], F. Carreón [c], S. Casanova [j], U. Cotti [e], J. Cotzomi [k], P. Desiati [l], E. De La Fuente [m], C. De León [e], N. Di Lalla [n], R. Diaz Hernandez [i], M.A. DuVernois [l], J.C. Díaz-Vélez [l], T. Ergin [g], C. Espinoza [a], N. Fraija [c], S. Fraija [c], J.A. García-González [o], F. Garfias [c], N. Ghosh [p], A. Gonzalez Muñoz [a], M.M. González [c,*], J.A. González [e], J.A. Goodman [q], J. Gyeong [r], J.P. Harding [s], I. Herzog [g], D. Huang [q], F. Hueyotl-Zahuantitla [b], A. Iriarte [c], S. Kaufmann [t], D. Kieda [u], A. Lara [v], W.H. Lee [c], J. Lee [w], H. León Vargas [a], J.T. Linnemann [g], A.L. Longinotti [c], G. Luis-Raya [t], K. Malone [s], O. Martinez [k], J. Martínez-Castro [x], H. Martínez-Huerta [y], J.A. Matthews [z], P. Miranda-Romagnoli [aa], P.E. Mirón-Enriquez [c], J.A. Morales-Soto [e], E. Moreno [k], M. Mostafá [f], M. Najafi [p], A. Nayerhoda [j], L. Nellen [ab], M.U. Nisa [g], R. Noriega-Papaqui [aa], L. Olivera-Nieto [ac,ad], N. Omodei [n], M. Osorio-Archila [c], E. Ponce [k], Y. Pérez Araujo [a], E.G. Pérez-Pérez [t], A. Pratts [a,*], C.D. Rho [r], A. Rodriguez Parra [e], D. Rosa-González [i], M. Roth [s], D. Salazar-Gallegos [g], A. Sandoval [a], M. Schneider [q], J. Serna-Franco [a,*], A.J. Smith [q], Y. Son [w], R.W. Springer [u], O. Tibolla [t], K. Tollefson [g], I. Torres [i], R. Torres-Escobedo [ae], E. Varela [k], L. Villaseñor [k], X. Wang [af], Z. Wang [af], I.J. Watson [w], H. Wu [l], S. Yu [ag], X. Zhang [j], H. Zhou [ae]

[a] *Instituto de Física, Universidad Nacional Autónoma de México, Ciudad de Mexico, Mexico*
[b] *Universidad Autónoma de Chiapas, Tuxtla Gutiérrez, Chiapas, Mexico*
[c] *Instituto de Astronomía, Universidad Nacional Autónoma de México, Ciudad de Mexico, Mexico*

* Corresponding authors.

*E-mail addresses:* ruben@fisica.unam.mx (R. Alfaro), crabpulsar@hotmail.com (C. Alvarez), alexis.andres1210@gmail.com (A. Andrés), earangel@astro.unam.mx (E. Anita-Rangel), miguel.araya@ucr.ac.cr (M. ), juan.arteaga@umich.mx (J.C. Arteaga-Velázquez), doavila@astro.unam.mx (D. Avila Rojas), hugo.ayala.solares@temple.edu (H.A. ), baburish@msu.edu (R. ), priyadarshini.bangale@temple.edu (P. ), belmont@fisica.unam.mx (E. ), abel@astro.unam.mx (A. Bernal), karen.scm@gmail.com (K.S. ), tcapistranc@gmail.com (T. ), alberto@inaoep.mx (A. ), mfcarreon@astro.unam.mx (F. Carreón), sabrinacasanova@gmail.com (S. ), umberto.cotti@umich.mx (U. ), jcotzomi@yahoo.com.mx (J. ), paolo.desiati@icecube.wisc.edu (P. Desiati), edfuente@gmail.com (E. ), cederik.de.leon@umich.mx (C. ), niccolo.dilalla@stanford.edu (N. Di Lalla), dihera77@gmail.com (R. Diaz Hernandez), duvernois@icecube.wisc.edu (M.A. ), juancarlos@icecube.wisc.edu (J.C. ), ergintul@msu.edu (T. ), m.catalina@fisica.unam.mx (C. ), nifraija@astro.unam.mx (N. ), sarafraija@hotmail.com (S. Fraija), anteus79@tec.mx (J.A. ), fergar@astro.unam.mx (F. ), nghosh1@mtu.edu (N. Ghosh), adiv.gonzalez@itoaxaca.edu.mx (A. Gonzalez Muñoz), magda@astro.unam.mx (M.M. González), anteus79@tec.mx (J.A. González), goodman@umd.edu (J.A. ), kyoungjh1011@naver.com (J. Gyeong), jpharding@lanl.gov (J.P. ), herzogia@msu.edu (I. ), dezhih@mtu.edu (D. ), filihz@gmail.com (F. ), airiarte@astro.unam.mx (A. ), skaufmann13@googlemail.com (S. Kaufmann), dave.kieda@utah.edu (D. ), alara@igeofisica.unam.mx (A. ), wlee@astro.unam.mx (W.H. ), jason.lee@uos.ac.kr (J. ), hleonvar@fisica.unam.mx (H. ), alonginotti@astro.unam.mx (A.L. ), gilura6969@hotmail.com (G. ), kmalone@lanl.gov (K. ), omartin@fcfm.buap.mx (O. ), macj@cic.ipn.mx (J. ), humberto.martinezhuerta@udem.edu (H. ), johnm@unm.edu (J.A. ), pa.miranda.r@gmail.com (P. ), pelimi92@gmail.com (P.E. ), jmoralessg@gmail.com (J.A. ), emoreno@fcfm.buap.mx (E. ), miguel@psu.edu (M. ), mnajafi@mtu.edu (M. Najafi), amid.nayerhoda@gmail.com (A. Nayerhoda), lukas@nucleares.unam.mx (L. ), nisamehr@msu.edu (M.U. ), ropapaqui@gmail.com (R. ), Laura.Olivera-Nieto@mpi-hd.mpg.de (L. ), nicola.omodei@stanford.edu (N. ), mosorio@astro.unam.mx (M. Osorio-Archila), eponce@fcfm.buap.mx (E. Ponce), yuniorpy@gmail.com (Y. ), egperezp@yahoo.com.mx (E.G. ), yoba_m_t_a@ciencias.unam.mx (A. Pratts), no397@naver.com (C.D. ), ancelmo.rodriguez@umich.mx (A. Rodriguez Parra), danrosa@inaoep.mx (D. ), mattroth@lanl.gov (M. Roth), salaza82@msu.edu (D. Salazar-Gallegos), asandoval@fisica.unam.mx (A. ), mschnei4@umd.edu (M. ), j.serna.1308@gmail.com (J. Serna-Franco), asmith8@umd.edu (A.J. ), youngwan.son@cern.ch (Y. Son), wayne.springer@utah.edu (R.W. ), ...tibolla@gmail.com (O. Tibolla), ...d.edu (Z. Wang), ian.jame... tollefson@pa.msu.edu (K. ), ibrahim.torres23@gmail.com (I. ), torresramiro350@sjtu.edu.cn (S. Yu), (R. ), ...zhangxx...@...com (X. Zhang), (E. Varela), ...@sjtu.edu.cn (H. ).

[d] Universidad de Costa Rica, San José 2060, Costa Rica
[e] Universidad Michoacana de San Nicolás de Hidalgo, Morelia, Mexico
[f] Department of Physics, Temple University, 1925N. 12th Street, 19122, Philadelphia, PA, USA
[g] Department of Physics and Astronomy, Michigan State University, East Lansing, MI, USA
[h] Universitá degli Studi di Torino, I-10125, Torino, Italy
[i] Instituto Nacional de Astrofísica, Óptica y Electrónica, Puebla, Mexico
[j] Institute of Nuclear Physics, Polish Academy of Sciences, PL-31342, IFJ-PAN, Krakow, Poland
[k] Facultad de Ciencias Físico Matemáticas, Benemérita Universidad Autónoma de Puebla, Puebla, Mexico
[l] Dept. of Physics and Wisconsin IceCube Particle Astrophysics Center, University of Wisconsin—Madison, Madison, WI, USA
[m] Departamento de Física, Centro Universitario de Ciencias Exactase Ingenierias, Universidad de Guadalajara, Guadalajara, Mexico
[n] Department of Physics, Stanford University, 94305-4060, Stanford, CA, USA
[o] Tecnologico de Monterrey, Escuela de Ingeniería y Ciencias, Ave. Eugenio Garza Sada 2501, 64849, Monterrey, N.L, Mexico
[p] Department of Physics, Michigan Technological University, Houghton, MI, USA
[q] Department of Physics, University of Maryland, College Park, MD, USA
[r] Department of Physics, Sungkyunkwan University, 16419, Suwon, South Korea
[s] Los Alamos National Laboratory, Los Alamos, NM, USA
[t] Universidad Politecnica de Pachuca, Pachuca, Hgo, Mexico
[u] Department of Physics and Astronomy, University of Utah, Salt Lake City, UT, USA
[v] Instituto de Geofísica, Universidad Nacional Autónoma de México, Ciudad de Mexico, Mexico
[w] University of Seoul, Seoul, South Korea
[x] Centro de Investigación en Computación, Instituto Politécnico Nacional, México city, Mexico
[y] Departamento de Física y Matemáticas, Universidad de Monterrey, Monterrey, NL, Mexico
[z] Dept of Physics and Astronomy, University of New Mexico, Albuquerque, NM, USA
[aa] Universidad Autónoma del Estado de Hidalgo, Pachuca, Mexico
[ab] Instituto de Ciencias Nucleares, Universidad Nacional Autónoma de Mexico, Ciudad de Mexico, Mexico
[ac] Max-Planck Institute for Nuclear Physics, 69117, Heidelberg, Germany
[ad] GRAPPA & Anton Pannekoek Institute for Astronomy, University of Amsterdam, Science Park 904, 1098 XH, Amsterdam, the Netherlands
[ae] Tsung-Dao Lee Institute & School of Physics & Astronomy, Shanghai Jiao Tong University, 800 Dongchuan Rd, 200240, Shanghai, SH, China
[af] Department of Physics, Missouri University of Science and Technology, Rolla, MO, USA
[ag] Department of Physics, Pennsylvania State University, University Park, PA, USA

ARTICLE INFO

*Keywords:*
Dark matter
Axion-like particles
Astroparticle physics
Gamma-ray observatories
Extragalactic astronomy
Relativistic jets

ABSTRACT

In this work, we perform an indirect search for axion-like particles (ALPs) through their hypothesized mixing with photons in the presence of magnetic fields. ALPs are a well-motivated dark-matter candidate class, and the photon-ALP conversion mechanism provides a unique channel to constrain their mass and coupling constant using very-high-energy gamma-ray observations. The photon-ALP mixing could alter the observed gamma-ray spectrum from extragalactic sources by effectively reducing the apparent attenuation due to extragalactic-background-light absorption. We analyze 7.5 years of data from the High Altitude Water Cherenkov (HAWC) Observatory, targeting the nearby radio galaxy M87. This source is located within the Virgo cluster and is an ideal environment for photon-ALP conversion due to its low redshift and the large-scale, strongly magnetized medium of the cluster. We find no evidence for a photon-ALP conversion signal and, consequently, set constraints on the ALP mass and photon-ALP coupling constant with emission from M87 which are consistent with previous results. Our analysis places competitive constraints on the ALP parameter space, defining an exclusion region in the mass range of approximately $10^{-8}$ to $10^{-6}$ eV for coupling constants above $5 \times 10^{-12}$ GeV$^{-1}$, complementing previous constraints from other gamma-ray observatories.

## 1. Introduction

The existence of dark matter remains one of the most significant open questions in modern physics and cosmology, motivating searches for physics Beyond the Standard Model (BSM). The standard cosmological model, ΛCDM, posits that dark matter accounts for approximately 27% of the total matter density in the Universe. While its gravitational effects are well-established, its particle nature is yet to be confirmed [1,2].

Among the many theoretical candidates beyond the Standard Model of particle physics, Axions, and more generally Axion-Like Particles (ALPs), are compelling candidates for dark matter. The theoretical basis stems from the Peccei-Quinn mechanism, where the Axion is proposed to resolve the strong Charge-Parity (CP) problem in Quantum Chromodynamics (QCD) [3–5]. ALPs, however, are a broader class of hypothetical particles that extend this idea without necessarily solving the CP problem[6].

ALPs are predicted to have masses from extremely light ($10^{-24}$ eV) [7,8] to heavy ($MeV$s or more) [9,10]. Direct detection of these particles in laboratory experiments is particularly challenging due to their extremely weak coupling with Standard Model particles, which is typically suppressed by a high-energy scale $f_a$ [3,4]. This results in exceptionally small interaction cross-sections, requiring either highly sensitive resonant detectors that only probe narrow mass windows or extremely large detector volumes to identify a signal above background noise.

The search for ALPs is particularly promising in the context of high-energy astrophysics, where the interaction of ALPs with photons can be probed through a phenomenon known as photon-ALP conversion. This process occurs in the presence of strong magnetic fields and can lead to observable distortions in the energy spectra of high-energy gamma-ray sources [11]. Specifically, ALPs can travel great distances without significant absorption, whereas high-energy photons are subject to attenuation by the Extragalactic Background Light (EBL) [12–15]. If photons from a distant source convert into ALPs near the source and then back into photons closer to Earth, the observed spectrum will show a higher flux than predicted by EBL absorption alone [6,16], an effect known as *photon resurrection*. Furthermore, because the photon-ALP conversion probability depends on the energy of the photon and the structure of the intervening magnetic fields, the mixing process typically induces stochastic "spectral wiggles" –irregularities or fluctuations in the energy spectrum that cannot be explained by standard astrophysical power laws [17–19]. These spectral distortions are a hallmark of ALP interactions in turbulent environments like galaxy clusters.

The search for ALP-photon mixing requires a source with a well-understood, high-energy gamma-ray spectrum and a suitable magnetic field environment along the line of sight. The giant radio galaxy M87, at a distance of $16.7 \pm 0.2$ Mpc [20], presents an ideal case for such an investigation. Unlike many extragalactic gamma-ray sources, which are typically blazars with highly variable emission, M87 exhibits less pro-nounced amplitude and slower timescales of flux variations over time [21,22]. This stability is crucial for a time-integrating observatory like the High Altitude Water Cherenkov (HAWC) observatory, enabling deep and statistically significant measurements of its time-averaged spectrum.

Located at a distance of $16.7 \pm 0.2$ Mpc [20], M87 is embedded in the Virgo cluster, providing an extensive path length for photon-ALP oscillations. The conversion occurs across two primary regions: the high-field environment of the M87 jet ($B \sim 10 - 100\ \mu$G) extending over several kiloparsecs [23], and the intra-cluster medium (ICM), which maintains a magnetized environment out to a virial radius of approximately 1.5-2 Mpc [24,25].

The conversion efficiency over these scales is governed by the oscillation length:

$$L_{\rm osc} = \frac{4\pi E}{m_a^2} \lesssim L_{\rm coh}. \tag{1}$$

To avoid destructive phase-averaging, $L_{osc}$ must be comparable to or larger than the magnetic field coherence length ($L_{coh}$). To quantify the spatial scale required for efficient mixing, our simulations assume a conservative magnetic field strength of 3 $\mu$G within each coherence domain-a value typical of general galaxy clusters. Even under this conservative assumption, we find that the effective conversion distance ($L_{\rm max}$) is sufficiently short. At 10 TeV, the required distance varies from approximately 33 kpc for $m_a = 5 \times 10^{-8}$ eV to less than 2 kpc for $m_a = 5 \times 10^{-7}$ eV. These scales are well within the extensive magnetized environment of the Virgo Cluster, which, characterized by a significantly stronger field of 34.2 $\mu$G at its core [26], provides a path length far exceeding the requirements for efficient conversion in the $10^{-9} - 10^{-7}$ eV mass range [6,11].

While blazars like Mrk 421 or PKS 2155-304 are strong gamma-ray emitters, their jets often point directly toward us, which limits the observed interaction path length with their host galaxies' magnetic fields. In contrast, the moderate inclination of M87's jet provides an advantageous geometry for studying magnetic-field effects along the line of sight. This comprehensive magnetic environment offers a distinct advantage for probing photon-ALP conversions over a longer, highly magnetized path length. Furthermore, this source is particularly advantageous because the intracluster magnetic fields of other host clusters are often less characterized or significantly weaker than the Virgo Cluster's [27,28]. While clusters such as Perseus or Coma have been extensively studied, the vast majority of galaxy clusters lack the detailed Faraday rotation measure maps or high-resolution X-ray data necessary to constrain their magnetic field profiles with the precision available for Virgo [24,25,27].

M87 has been widely studied across the electromagnetic spectrum. Its complex jet structure and core emission have been extensively mapped in the radio band by the Event Horizon Telescope and VLBA [29–31], in the optical by the Hubble Space Telescope [32], and in X-rays by Chandra and XMM-Newton [33,34], as well as multiwavelength studies [35,36]. With a redshift of $z = 0.0044$ [20], it is the closest radiogalaxy within HAWC's field-of-view (FoV) and the second closest observed radio galaxy to date.

Moreover, M87 hosts a supermassive black hole with a mass of approximately $(6.5 \pm 0.7) \times 10^9\ M_\odot$ [29] and a relativistic jet which extends between 1.5 and 2 kpc, with an estimated angle of 15° to 25° relative to the line of sight [37]. The gamma-ray emission from M87's jet has been studied extensively by Cherenkov telescopes such as HEGRA [38], H.E.S.S. [39,40], MAGIC [41–43], and VERITAS [44]. While earlier studies using wide FoV instruments like the HAWC Observatory [45,46] reported only marginal detections below the $5\sigma$ threshold [47,48], the situation has evolved with increased exposure and improved reconstruction techniques. Currently, wide-field observatories have firmly established M87 as a VHE source; HAWC [21] and LHAASO [22] have detected it with $> 5\sigma$ up to ∼20 TeV. In addition to its detection, most of the Virgo cluster's mass-to-light ratio ($M/L$) is associated with M87 which has been confirmed with kinematic studies that have shown that its $M/L$ goes from $\mathcal{O}(10) \lesssim M_{M87}/L_{M87} \lesssim \mathcal{O}(100)$ [49–51].

The astrophysical conditions of the Virgo cluster and the recent detection of M87 emissions above 10 TeV by the HAWC observatory highlights its importance as a prime target for high-energy astrophysics and searches for exotic particles, such as ALPs. Previous analyses have been performed at different wavelengths including microwaves [52], infrared [53], and gamma-rays with H.E.S.S[26]. and LHAASO [54].These characteristics make M87's line of sight particularly promising for enhancing the ALP-photon conversion signal, then HAWC observations of M87 provide an opportunity to search for these photon-ALP mixing signatures.

In this paper, we present a dedicated search for photon-ALP conversions using a dataset exceeding seven years from the HAWC Observatory centered on M87. We detail our analysis methodology, present the resulting statistical limits on the ALP parameter space, and discuss the implications of our findings in the context of existing experimental constraints. The scope of this work focuses on the regime where the photon-ALP conversion mechanism is most effectively probed by high-energy astrophysical observations; specifically, we explore ALP masses spanning $m_a \sim 10^{-8}$ to $\sim 10^{-6}$ eV and coupling constants ranging from $g_{a\gamma\gamma} \sim 10^{-12}$ to $\sim 10^{-10}$ GeV$^{-1}$.

## 2. The HAWC observatory

The HAWC observatory is a ground-based gamma-ray and cosmic-ray detector located at an altitude of 4,100 m on the Sierra Negra volcano in Mexico [45,46]. It is an array of 300 large water Cherenkov detectors (WCDs) covering an area of approximately 22,000 m$^2$. Each WCD is a cylindrical detector containing 200,000 liters of purified water and equipped with four photomultiplier tubes (PMTs) at its bottom, designed to detect extensive air showers produced by gamma-rays and cosmic rays. When a high-energy gamma-ray or cosmic-ray particle interacts with the atmosphere, it produces a cascade of secondary particles, known as an extensive air shower. When these relativistic particles travel through the water in the WCDs, they produce Cherenkov light, which is then detected by the PMTs. The relative timing of the PMT signals is used to reconstruct the direction of the primary particle, while the amplitude of the signals is used to estimate its energy along a wide range that goes from 300 GeV up to 100 TeV. HAWC's wide instantaneous FoV of $\sim 2$ sr allows it to observe the sky overhead at any given moment. As the Earth rotates, this enables the observatory to monitor approximately two-thirds of the entire sky during a 24-hour period. This extensive sky coverage, combined with a high duty cycle (> 95%) allow for continuous monitoring of the northern sky, providing a large dataset for sources like M87 that transit overhead. This capability is crucial for detecting subtle spectral distortions that might be present only at the highest energies.

To improve its sensitivity at the highest energies, HAWC has been upgraded with a sparse outrigger array. The outrigger array consists of 350 smaller WCDs [55], each with a diameter of 1.55 m and a height of 1.65 m, equipped with a single PMT. These outriggers are sparsely distributed around the main array, increasing the instrumented area by a factor of 4 to 5. By constraining the location of the shower core, the outrigger array resolves ambiguities in shower reconstruction and increases the number of well-reconstructed showers above multi-TeV energies, which is critical for enhancing HAWC's sensitivity at the highest energies. It is important to note that the data utilized in this analysis were collected exclusively from the main array, as the outrigger array was not fully integrated into the data-processing chain for the 7.5 years period under study.

Key performance characteristics of HAWC relevant to this analysis include its wide FoV, which allows M87 to be observed for approximately 6 hours per day, and its high duty cycle, ensuring consistent data collection. The observatory has an angular resolution that improves from approximately 1.0° at 300 GeV to better than 0.2° for energies above 10

TeV [48]. Furthermore, its effective area increases with energy, rising from $\sim 10^2$ m$^2$ at 300 GeV to a plateau of approximately $10^5$ m$^2$ for energies exceeding 10 TeV [48]. Another instrumental parameter crucial for the detection of spectral wiggles is the HAWC's energy resolution, which is approximately $\lesssim 40\%$ at 1 TeV and improves to better than 23% for energies above 10 TeV, assuming the Neural-Networks (NN)-based energy estimator [56]. This precision in energy reconstruction is vital for resolving the energy-dependent spectral distortions. These combined characteristics make HAWC particularly sensitive to gamma rays in the multi-TeV range, which is essential for detecting the subtle spectral distortions predicted from photon-ALP mixing.

The gamma/hadron separation is performed using a "compactness" ($C$) parameter that characterizes the lateral distribution of the air shower on the ground. This parameter distinguishes the relatively smooth energy profiles of gamma-ray-induced showers from the more irregular and "clumpy" profiles typical of hadronic cosmic-ray backgrounds. A detailed description of the parameter's construction and its performance within the HAWC reconstruction pipeline is provided in Refs. [45,48,56].

## 3. Data analysis & methodology

Raw data from the HAWC detector are processed to reconstruct the direction and energy of the primary gamma-ray. The core of the air shower is reconstructed based on the pattern of hit PMTs. The direction of the primary particle is then determined from the arrival times of the Cherenkov photons at each WCD. The primary energy is estimated, based on the total charge deposited in the PMTs, with the NN-based energy estimator[56] to improve energy resolution at higher energies. A crucial step in the analysis is the rejection of the overwhelming background of cosmic-ray induced air showers, which is achieved by applying event selection criteria and a gamma/hadron separation parameter [48].

The analysis presented here utilizes a total of 7.5 years of HAWC data. The target region of interest is centered on the known position of M87, presenting a maximum significance of $6.1\sigma$ of signal over the background. The peak of this emission is found at a distance $< 1^\circ$ from the reported position of M87, which is fully consistent with the instrument's Point Spread Function (PSF) at the energy threshold of $\sim 1$ TeV. Specifically, for a source at M87's declination, the HAWC angular resolution at 1 TeV is approximately $0.8^\circ$ [48]. As shown in Fig. 1 left, while the angular resolution improves to $\sim 0.2^\circ$ at higher energies ($> 10$ TeV), the significance map is dominated by the higher statistics available at lower energies where the PSF is broader.

Significance map (Fig. 1, right) shows several point-like excesses in the vicinity, though these are all of low significance (below the $5\sigma$ discovery threshold) and are consistent with statistical fluctuations of the background. Since M87 is modeled as a point source, the spectral analysis is localized to its known coordinates. Above 1 TeV the PSF is narrow enough to ensure that the analysis is not significantly contaminated by the faint, nearby fluctuations visible in the significance map (Fig. 1, right). This spatial isolation allows us to consider that the derived spectrum for M87 is a robust representation of its intrinsic emission.

For the spectral analysis of M87, we followed the methodology previously described in [21], which includes a detailed cross-calibration and comparison of HAWC's spectral reconstruction with results from IACTs like H.E.S.S, MAGIC, and VERITAS. By adopting this validated framework, we compared our 7.5-year integrated spectral parameters with other observations of M87 presented in [57], ensuring that our dataset corresponds to an intermediate activity state of the source. In this work, we assumed M87 to be a point-like gamma-ray source, as its angular size is smaller than the PSF of the HAWC observatory in the relevant energy range, using the `Gammapy` framework [58]. We modeled the intrinsic gamma-ray emission of M87 using a Power Law spectral distribution. This spectrum was then convolved with the Finke et al. (2010) [15] and Franceschini et al. (2008) [12] EBL attenuation models in parallel, performing an independent fit for each EBL model leading to the parameters shown in Table 1, and their energy spectrum distributions are shown in Fig. 1 (left). Specifically, the spectral index and flux normalization derived here are totally consistent with the published results from the 5-year M87 HAWC analysis [21] and are also consistent with recent LHAASO observations of the source [22], confirming the robustness of our data selection and fitting methodology.

**Table 1**
Power Law fit parameters estimated for M87's energy spectrum, considering both Franceschini et al. (2008) [12] and Finke et al. (2010) [15] EBL models. The errors presented are associated to the statistical uncertainties.

| EBL Model | Normalization at 1 TeV $\times 10^{-13}$ TeV$^{-1}$ s$^{-1}$ cm$^{-2}$ | Spectral index |
|---|---|---|
| Finke et al. (2010) [15] | $4.40 \pm 1.81$ | $2.51 \pm 0.18$ |
| Franceschini et al. (2008) [12] | $3.52 \pm 0.72$ | $2.35 \pm 0.06$ |

The assumed EBL model is a significant source of uncertainty, as the photon-ALP conversion probability depends on the opacity of the extragalactic medium. We evaluate our results using the Franceschini et al. (2008) [12] and Finke et al. (2010) [15] models, which constitute the representative envelope of EBL density. The range of attenuation parameters (specifically the optical depth, $\tau$) is characterized by the underlying physical assumptions of the EBL models regarding the Cosmic Star Formation History (CSFH) and the Initial Mass Function (IMF). For instance, the Finke et al. (2010) model is characterized by a "forward-modeling" approach that calculates attenuation based on stellar population synthesis and dust re-emission. In contrast, the Franceschini et al. (2008) model is an empirical "backward-modeling" approach characterized by observed galaxy luminosity functions. The resulting opacity is strongly dependent on the energy of the gamma-ray photon ($E_\gamma$), as the pair-production cross-section peaks when the interaction with EBL photons of energy $\epsilon$ satisfies the threshold condition $\epsilon \cdot E_\gamma \approx 2\left(m_e c^2\right)^2$ [12,15]. The variation between these models represents the current systematic uncertainty in the infrared photon density, which is the primary factor determining the opacity for multi-TeV gamma rays from a nearby source like M87.

Since more recent models, such as Franceschini et al. (2017) [13], Domínguez et al. (2011) [14], or Saldana-Lopez et al. (2021) [59] provide $\tau$ values that fall within the range determined by the Franceshini et al. (2008) [12] and the Finke et al. (2010) [15] models, these two benchmarks effectively bracket the systematic uncertainties in EBL-induced opacity. Utilizing this range ensures that our ALP exclusion limits are robust against varying theoretical realizations of the EBL.

In addition to the extragalactic environment, another source of uncertainty is the magnetic field model, which governs the ALP-photon oscillation within and around the source. The parameters selected are based on best-fit values derived from observational literature (e.g., [23,26]). The specific configuration used in this study –including the jet's distribution model and spatial characteristic parameters– is detailed in Table 2, which provides a comprehensive list of the physical parameters and their corresponding literature sources.

To account for these physical and systematic variables, the data was analyzed using the `Gammapy v1.3` [58] and `3ML` [60] frameworks in parallel, whose performances are fully compatible as shown in [61], ensuring that our numerical results are independent of the specific software implementation.

The impact of these modeling choices is evident in our spectral results. The fitted spectra shown in Fig. 1 (left) provide a direct comparison between the two EBL models, Franceschini et al. (2008) [12] and Finke et al. (2010) [15], used in this work. Both models predict nearly identical attenuation up to approximately 10 TeV, after which a slight divergence appears, with the Franceschini et al. (2008) [12] model predicting stronger overall attenuation at higher energies. This difference, which represents a systematic uncertainty in the EBL, is fully considered in our final analysis and the determination of the ALP exclusion limits. Both EBL models provide an acceptable fit to the HAWC data points, indicating that our current level of sensitivity is insufficient to prefer one model over the other, thereby requiring us to consider both in assessing the systematic uncertainties of the analysis.

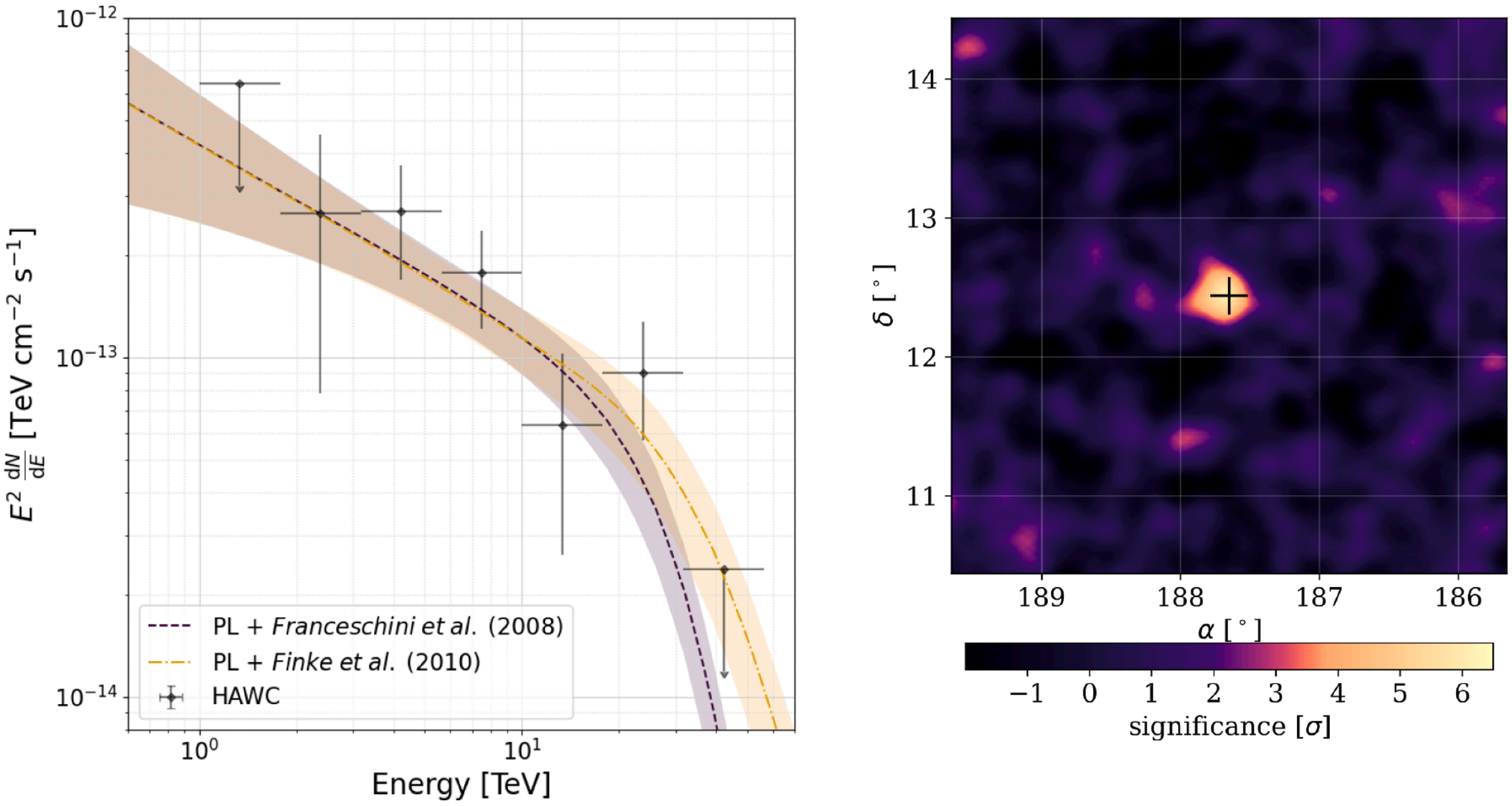

**Fig. 1. Left:** Energy spectrum of M87 assuming Franceschini et al. (2008) [12] and Finke et al. (2010) [15] EBL models. The upper limits were computed with a confidence level of 95%. **Right:** Significance map of M87 analyzing the 7.5 years of HAWC's data with a 6.1$\sigma$ emission.

**Table 2**
Parameters used to model the astrophysical environments that interact with the gamma-rays/ALPs, defined in the `gammaALPs` framework. The acronym SMBH stands for Super-Massive Black Hole, and the HST-1 is the first bright knot in the Jet of M87.

| Parameter | Value | Reference |
|---|---|---|
| Position (RA, Dec) | (187.69°, 12.39°) | [68] |
| Redshift ($z$) | 0.0044 | [20] |
| Angle of the jet with the l.o.s. ($\theta$) | 17° ± 8° | [69] |
| Lorentz factor ($\Gamma$) | 1.2 ± 0.1 | [69] |
| Jet's magnetic field ($B_{0,\mathrm{Jet}}$) | $6 \times 10^{-4}$ G | [23] |
| Distance from the SMBH to the HST-1 ($r_0$) | 70 pc | [66] |
| Longitude of the Jet ($R_{\mathrm{Jet}}$) | 2200 pc | [70] |
| Virgo's magnetic field ($B_{0,\mathrm{Virgo}}$) | 34.2 $\mu$G | [26] |
| Virgo's Abell radius ($r_{\mathrm{Abell}}$) | 1700 ± 200 kpc | [71] |
| EBL model | Franceschini et al. (2008) | [12] |
| | Finke et al. (2010) | [15] |
| Jet model | Helical and tangled model | [72] |
| Milky Way magnetic field model | Coherent component model | [65] |

## 4. ALPs survival probabilities

The hypothesized photon-ALP mixing, whereby a gamma-ray photon converts into an ALP in the presence of an external magnetic field, is formally referred to as the Primakoff effect [6]. This interaction between photons and ALPs is described by an extension of the Standard Model Lagrangian, which includes a term for the coupling between ALPs and photons. The full Lagrangian is given by:

$$\mathcal{L} = \mathcal{L}_{a\gamma} + \mathcal{L}_{EH} + \mathcal{L}_a. \tag{2}$$

The coupling term, $\mathcal{L}_{a\gamma}$, is written as:

$$\mathcal{L}_{a\gamma} = -\frac{1}{4} g_{a\gamma\gamma} F_{\mu\nu} \tilde{F}^{\mu\nu} a. \tag{3}$$

Here, $F_{\mu\nu}$ and $\tilde{F}^{\mu\nu}$ are the electromagnetic field tensor and its dual, respectively, $a$ is the ALP field, and $g_{a\gamma\gamma}$ is the coupling constant [6,11,16].

The second term on the right-hand side of Eq. (2) corresponds to the Euler-Heisenberg correction Lagrangian, which describes photon propagation in the presence of a strong electromagnetic field [16]. The last term is the ALP Lagrangian, which depends on the ALP mass ($m_a$) and its kinetic energy:

$$\mathcal{L}_a = \frac{1}{2}\partial_\mu a \partial^\mu a - \frac{1}{2} m_a^2 a^2 \tag{4}$$

In the presence of an external magnetic field, this coupling allows for the oscillation of photons into ALPs. The propagation of a photon-ALP system can be described by a state vector containing the two transverse photon polarization states and the ALP state. The evolution of this system is governed by a mixing matrix, which takes into account the different physical effects along the propagation path, including vacuum birefringence, photon-ALP mixing, and plasma effects. The mixing factor, which describes the strength of the ALP-photon oscillation, is described as follows [11,16,62,63]:

$$\Delta_{a\gamma} \simeq 1.52 \times 10^{-2} \left( \frac{g_{a\gamma\gamma}}{10^{-11}\ \mathrm{GeV}^{-1}} \right) \left( \frac{B_T}{10^{-3}\ \mu\mathrm{G}} \right) \mathrm{Mpc}^{-1}, \tag{5}$$

where $B_T$ is the component of the magnetic field transverse to the photon's direction of motion.

The efficiency of the photon-ALP conversion, $P_{a\gamma}$, is critically dependent on the distance the photon-ALP beam traverses and the structure of the intervening magnetic field [64]. Specifically, the probability is maximized when the photon-ALP oscillation length, $L_{\mathrm{osc}}$, is comparable to the distance over which the magnetic field remains approximately constant in strength and direction. This distance is known as the magnetic coherence length, $L_{\mathrm{coh}}$. When $L_{\mathrm{osc}} \approx L_{\mathrm{coh}}$, the conversion is coherent, leading to high conversion probability. Conversely, when $L_{\mathrm{coh}}$ is much shorter than $L_{\mathrm{osc}}$, the magnetic field direction changes rapidly, which suppresses the conversion efficiency.

The parameters used to characterize the astrophysical environments for this analysis, as listed in Table 2, are crucial for accurately modeling the photon-ALP conversion process. While M87 is the closest AGN inside the HAWC's FoV, its strong magnetic fields, particularly within its jet and the host Virgo cluster, are significant enough to compensate for its relatively close distance, enhancing the overall ALP survival probability. The physical parameters used to model M87 and the Virgo cluster's magnetic fields are consistent with commonly accepted values for similar astrophysical systems. Furthermore, we consider two widely used EBL models, Finke et al. (2010) [15] and Franceschini et al. (2008) [12], to account for systematic uncertainties related to gamma-ray absorption.

The Milky Way magnetic field chosen for this analysis is modeled following the Jansson & Farrar (2012) [65] framework. This choice is based on the model's ability to provide a significantly improved fit to both the WMAP7 Galactic Synchrotron Emission and over 40,000 extragalactic rotation measures, which are the primary observational constraints for the large-scale Galactic magnetic field. Crucially for studies of out-of-plane sources like M87, this model includes a poloidal "X-field" component and an extended halo field –features that were missing in earlier models but are consistently observed in external edge–on galaxies. This model has been successfully adopted in recent similar works (e.g., [26]), establishing it as a current benchmark in the field.

It is important to note that while the magnetic field near the central supermassive black hole is expected to be significantly stronger, the gamma-ray emission in our analysis is assumed to originate from the jet's first node [23,66], located farther from the core. For this reason, we adopt the jet magnetic field as the dominant photon-ALP conversion region.

However, for a realistic astrophysical environment such as M87's jet and the Virgo cluster, the magnetic field is not homogeneous and contains turbulent components. Therefore, a simple analytical formula is not sufficient to describe the full interaction. To accurately compute the survival probabilities, we use the `gammaALPs` software [67], which numerically solves the full propagation equations including the Euler-Heisenberg QED vacuum birefringence correction ($\mathcal{L}_{EH}$) [16,67]. This framework considers a detailed model for each astrophysical environment the photons and ALPs traverse: the M87 jet, the Virgo cluster medium, the intergalactic medium, and the Milky Way magnetic field. This approach accounts for multiple magnetic domains, turbulence, and coherence effects, providing a robust calculation of the final survival probability of a gamma-ray photon from the source to the HAWC detector.

The final observed flux can then be related to the intrinsic source flux by the conversion probability as:

$$\left.\frac{\mathrm{d}\phi}{\mathrm{d}E}\right|_{\text{detected}} = \left(1 - P_{a\gamma}\right) \cdot \left.\frac{\mathrm{d}\phi}{\mathrm{d}E}\right|_{\text{source}} \cdot e^{-\tau(E,z)} \tag{6}$$

In this equation, $P_{a\gamma}$ is the total survival probability contribution from the ALP mixing channel. It is essential to note that the `gammaALPs` framework solves the full coupled system of propagation equations, where the EBL attenuation is explicitly included in the diagonal elements of the mixing matrix alongside the plasma and QED terms. Therefore, the resulting overall photon survival probability derived from the `gammaALPs` transfer matrix already includes the effect of EBL absorption, rendering the separate explicit $e^{-\tau(E,z)}$ factor conceptual. For the final likelihood analysis, the total survival probability as computed by the framework, is used.

The ranges for the ALP mass, $m_a$, and photon-ALP coupling constant, $g_{a\gamma\gamma}$ were selected based on the physical conditions most favorable to a detectable photon-ALP conversion signal. The chosen mass range ($10^{-8}$ to $10^{-6}$ eV) corresponds to the regime where the conversion makes the possible detection feasible over astronomical distances, allowing for a measurable effect on the high-energy gamma-ray spectrum, in particular in the 1-100 TeV energy range covered by HAWC. For higher masses, the coherence length becomes too short, making the conversion probability negligible over the distance from M87. The coupling constant range ($10^{-12}$ to $10^{-10}$ GeV$^{-1}$) was chosen as it represents the parameter space where the conversion probability is sufficiently large to produce a spectral distortion that can be constrained by HAWC's sensitivity, while also complementing or improving upon existing exclusion limits from other experiments.

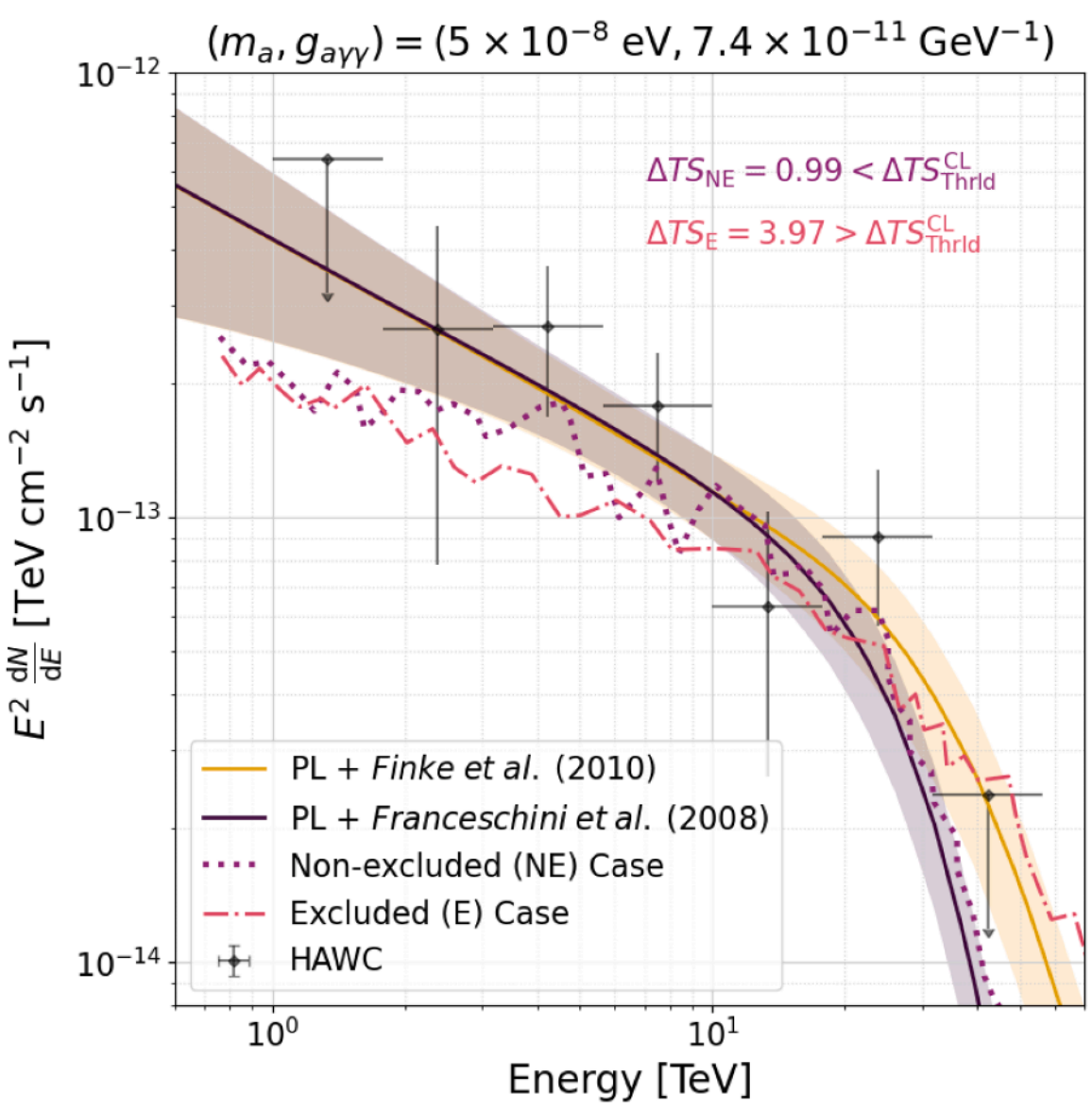

**Fig. 2.** Comparison of the HAWC observed spectrum of M87 with two representative realizations of a single ALP candidate $(m_a,\ g_{a\gamma\gamma}) = (5 \times 10^{-8}$ eV, $7.4 \times 10^{-11}$ GeV$^{-1})$. The curves illustrate how the stochasticity of the magnetic field –specifically the random orientation and polarization of the field components– leads to varying spectral outcomes for the same mass and coupling values. We classify a candidate realization as excluded only if the global $\Delta TS$ of the fit to the data exceeds the 95% quantile of the $\Delta TS$ distribution derived from null-hypothesis simulations (red dot-dashed curve). Conversely, a realization with a $\Delta TS$ below this threshold is considered non-excluded (violet dotted curve). These examples highlight that the detectability of an ALP signal is determined by a global statistical threshold rather than local visual compatibility with individual flux points. The solid lines represent the null hypothesis (only-EBL-attenuated spectrum) assuming the Franceschini et al. (2008) [12] (magenta solid line) and Finke et al. (2010) [15] (yellow solid line) EBL models. (For interpretation of the references to colour in this figure legend, the reader is referred to the web version of this article.)

The impact of magnetic field stochasticity on the predicted ALP signatures is illustrated in Fig. 2. Because the photon-ALP conversion probability is intrinsically sensitive to the specific orientation and polarization of the magnetized environment, different realizations of the same physical ALP candidate can produce markedly different spectral distortions. This inherent variability necessitates our statistical framework: we classify an ALP candidate as excluded only if the $\Delta TS$ value obtained from the global fit to the observed data is higher than the 95% quantile of the $\Delta TS$ distribution derived from our null-hypothesis simulations. Fig. 2 displays two distinct realizations for a single ALP hypothesis to highlight these potential outcomes. The non-excluded case (violet dotted curve) remains consistent with HAWC's flux points and 95% CL upper limits. In contrast, the excluded realization (red dot-dashed curve) arises from a magnetic field configuration that yields a $\Delta TS$ surpassing our statistical threshold, indicating a global incompatibility with the measured spectrum.

To account for the intrinsic uncertainty in the photon-ALP conversion probability, which arises from the pseudo-random initial polarization of the photons, we employ a statistical method that avoids introducing a priori assumptions. While some analyses, such as those in [73] and [74], rely on computing the mean or median of the survival probabilities from multiple simulations, this approach implicitly assumes a particular physical or mathematical representation of the average signal. The choice between using a mean or a median can directly and arbitrarily influence the shape and size of the fi-nal exclusion region, which lacks a strong theoretical or observational justification.

To overcome this limitation, our methodology follows the approach proposed in [75]. We perform 75 independent simulations for each ALP parameter pair ($m_a$,$g_{a\gamma\gamma}$), each with a randomly sampled initial photon polarization angle. This specific number was determined through prior convergence tests, ensuring that the calculated oscillation probability averaged to a stable, robust value while maintaining computational efficiency for the long-term data analysis. For each of these simulations, we compute a distinct exclusion region based on the resulting likelihood profile. The final, robust exclusion region is then defined as the intersection of all these individual exclusion regions, selecting the area that is excluded at a 95% CL in at least 95% of the simulations. This method bypasses the need to define a single, averaged survival probability. By instead comparing the individual exclusion regions and identifying their common, highly probable overlap, we arrive at a final constraint that is more robust and less susceptible to statistical and physical biases associated with mean- or median-based methods. This approach ensures that our derived constraints on the ALP parameter space are as conservative and hypothesis-independent as possible.

## 5. Statistical formalism

We search for the ALP signal using a binned Poisson likelihood-ratio test. The likelihood function is defined as:

$$\mathcal{L}(s, b|\text{data}) = \prod_{i=1}^{N_{bins}} \frac{(s_i + b_i)^{n_i}}{n_i!} e^{-(s_i+b_i)} \tag{7}$$

where $N_{bins}$ is the number of energy bins described in Table 3 of [48], $s_i$ and $b_i$ are the expected signal and background counts in bin $i$, and $n_i$ is the observed number of events in that bin, where the background counts were estimated using the direct-integration method applied to the source region, as described in [48].

The Test Statistic ($TS$) is defined as the ratio of the likelihood with a signal hypothesis to the likelihood with the background-only hypothesis, $TS = -2\log(\mathcal{L}_{s+b}/\mathcal{L}_b)$. The standard $\Delta TS \approx \chi^2$ relationship derived from Wilks' theorem [76] is not applicable in this analysis for several reasons. Primarily, the alternative hypothesis (including ALPs) is non-linear in the parameter space and represents a non-nested model relative to the null hypothesis (no ALPs), as the ALP physics fundamentally alters the photon propagation mechanism. Furthermore, the ALP survival probability exhibits degenerate solutions, making the likelihood function non-convex and highly complex [75,77,78]. These factors necessitate the use of the profile likelihood method, where the CL is determined by directly simulating the null hypothesis, as described in [79] to compute the $TS$ values to establish accurate CLs. To estimate the particular $TS$ values of this analysis, 5000 Poisson-fluctuated simulations were done using `Gammapy` as well, assuming the intrinsic energy spectra reported in Table 1 considering both EBL models. Fig. 3 presents the $\Delta TS$ distribution, where:

$$TS^{\text{Src}}_{\text{woALPs}} - TS^{\text{Src}}_{\text{wALPs}} = -2\log\left(\frac{\mathcal{L}_{\text{wALPs}}}{\mathcal{L}_{\text{woALPs}}}\right) = \Delta TS < \Delta TS^{\text{CL}}_{\text{Thrld}} \tag{8}$$

$TS^{\text{Src}}_{\text{woALPs}}$ is the $TS$ of the source without ALP contribution effects in the spectrum (the null hypothesis in this work) and $\mathcal{L}_{\text{woALPs}}$ is its associated likelihood function, $TS^{\text{Src}}_{\text{wALPs}}$ is the $TS$ of the source assuming ALP contribution effects in the spectrum (not-null hypothesis) and its related likelihood function $\mathcal{L}_{\text{wALPs}}$, and finally $\Delta TS^{\text{CL}}_{\text{Thrld}}$ is the $\Delta TS$ threshold for the chosen CL. Table 3 presents the corresponding $\Delta TS$ values for CLs between $1\sigma$ and $3\sigma$.

## 6. Results and discussion

From the search for a spectral anomaly indicative of photon-ALP conversion, the likelihood analysis yields no significant evidence for a photon-ALP conversion signal in the 7.5 years of M87 data detected with HAWC. The maximum test statistic achieved was below the threshold required for a 95% CL detection. Consequently, we proceed to calculate exclusion regions in the ALP mass-coupling constant parameter space ($m_a, g_{a\gamma\gamma}$) that are consistent with previous astrophysical constraints.

**Table 3**
$\Delta TS$ values for each CL from $1\sigma$ to $3\sigma$, assuming both Franceschini et al. (2008) [12] and Finke et al. (2010) [15] EBL models.

| Sigmas [$\sigma$] | CL [%] | $\Delta TS$ Franceschini et al. (2008) [12] | $\Delta TS$ Finke et al. (2010) [15] |
|---|---|---|---|
| 1 | 68.0 | 0.87 | 0.88 |
| 2 | 95.0 | 3.64 | 3.75 |
| 3 | 99.7 | 6.59 | 6.85 |

The exclusion regions at the 95% CL, considering the respective $\Delta TS$ values presented in Table 3, are shown in Fig. 4. There, we present a comparison of two distinct statistical treatments to account for the stochastic nature of the magnetic field realizations. The first, labeled as the 95% Ensemble Coverage Method (the methodology presented in this work), is derived by analyzing each individual simulation and determining the parameter space excluded by 95% of the realizations. This provides a comprehensive view of the systematic uncertainties inherent in the magnetic field modeling. The second approach, the Mean Survival Probability Method, simplifies the ensemble by using the average survival probability to define a single representative limit. As shown in Fig. 4, while the peak sensitivity remains centered at $m_a \approx 10^{-7}$ eV for both techniques, the mean-probability method provides a more localized exclusion region that characterizes the average response of the system, whereas the robust analysis accounts for the full variance of the magnetic field realizations.

To contextualize these findings within the current state of ALP searches, Fig. 5 compares our robust exclusion regions with constraints derived from other gamma-ray observatories. Our results for M87 complement existing limits from objects such as NGC 1275 [80] and radio-Quasars [81] both with Fermi-LAT, PKS 2155-304 with H.E.S.S. [82], and the Perseus cluster observed with MAGIC [83], among others, particularly by extending the exclusion sensitivity in the $10^{-7}$ eV mass range where the large-scale magnetic fields of the Virgo cluster facilitate efficient conversion. Our results provide valuable independent validation that helps map the viable parameter space for ALPs, though substantial regions remain unexplored. ALPs could still constitute dark matter in this mass range if their photon coupling is below our detection threshold.

These are generated by profiling the likelihood function over the nuisance parameters for both the Franceschini et al. (2008) [12] and Finke et al. (2010) [15] EBL model hypotheses. The similarity between the results for the two EBL models is a direct consequence of M87's low redshift ($z = 0.0044$). At this distance of approximately 16.7 Mpc, the gamma-ray path length is relatively short, limiting the cumulative effect of EBL attenuation. While the models predict increasing divergence in attenuation above 10 TeV, the resulting exclusion limits differ by less than 3% across the tested parameter space. Since neither EBL model can be statistically excluded by our current data, we adopt the Finke et al. (2010) [15] model results as our primary constraint, as this model predicts weaker EBL attenuation and therefore yields more conservative ALP limits that remain valid across the full range of EBL systematic uncertainty.

The overlap of our limits with existing constraints, despite 7.5 years of continuous observation, can be attributed to several factors. Systematic uncertainties in the magnetic field parameters contribute significantly to the overall uncertainty budget, with variations in both the Virgo cluster field strength and jet magnetic field introducing substantial uncertainties in the predicted conversion probabilities. Additionally, while M87 is detected at $6.1\sigma$ significance in the integrated count map, the spectral measurement uncertainties, particularly at the highest ener-gies where ALP signatures would appear, limit our sensitivity to subtle spectral distortions. Furthermore, our conservative approach of using the intersection of individual exclusion regions from 75 independent polarization realizations ensures robustness but may prevent us from achieving more restrictive limits.

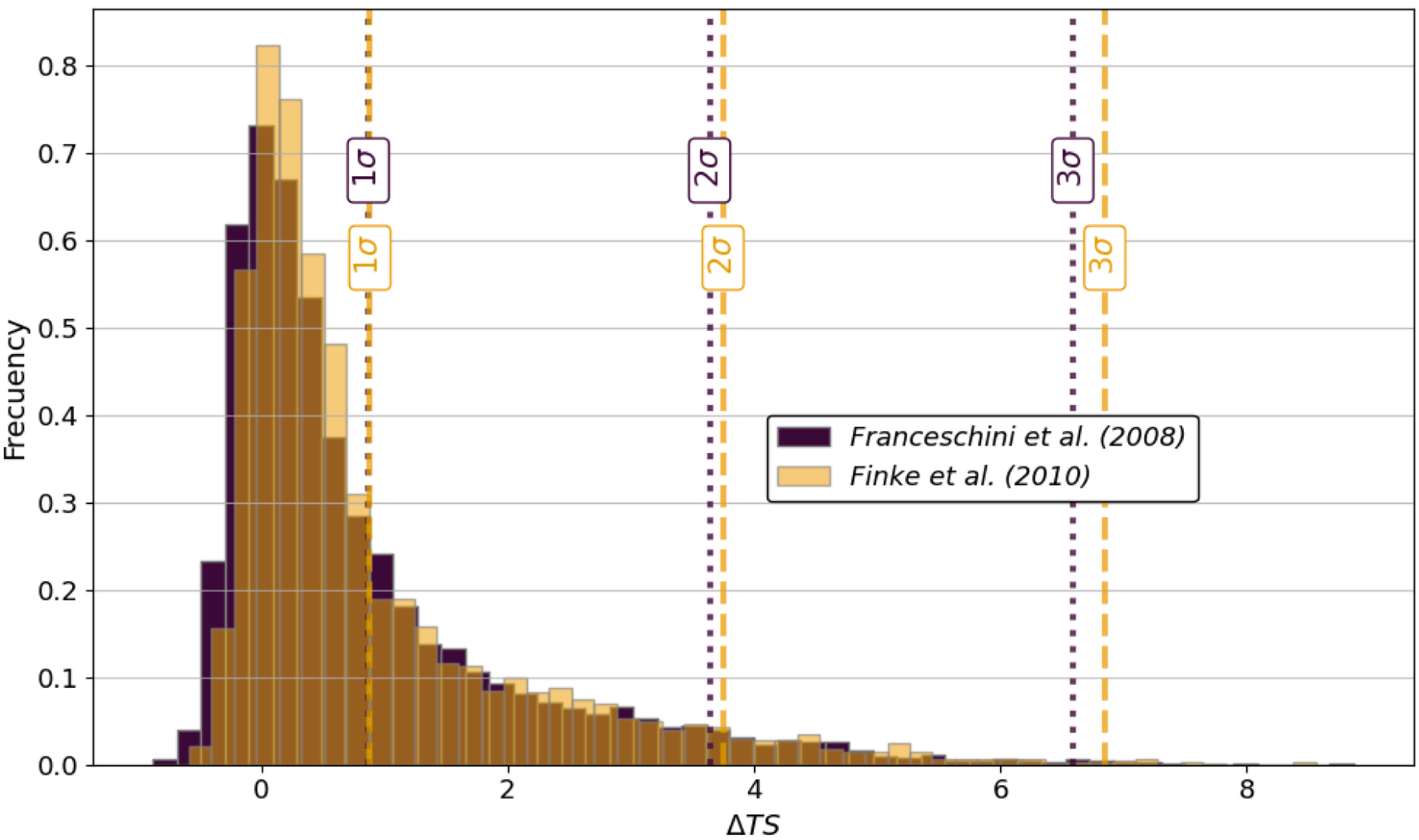

**Fig. 3.** Distributions of the max ($\Delta TS$) values simulating M87 emission observed with HAWC, and assuming the Franceschini et al. (2008) [12] (magenta) and Finke et al. (2010) [15] (dark yellow) EBL models. The vertical lines show the $\Delta TS_{\text{Thrld}}$ values for quantiles from $1\sigma$ to $3\sigma$, for the Franceschini et al. (2008) [12] and Finke et al. (2010) [15] EBL models, in magenta-dotted and yellow-dashed lines respectively. (For interpretation of the references to colour in this figure legend, the reader is referred to the web version of this article.)

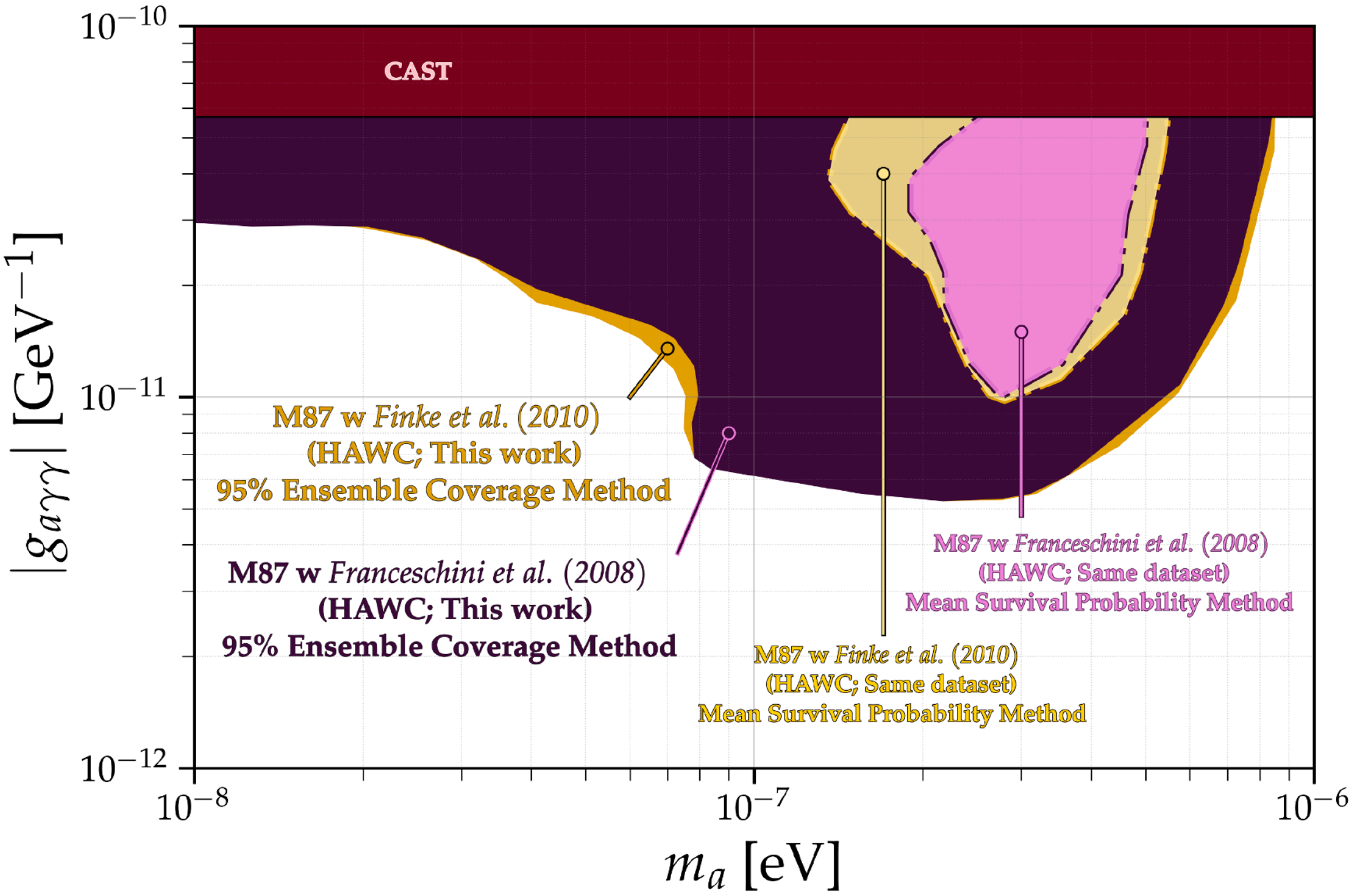

**Fig. 4.** 95% C.L. exclusion regions for the ALP-photon coupling $g_{a\gamma\gamma}$ vs. ALP mass $m_a$. The dark regions represent the robust limits for different EBL models following the statistical method presented in this work, while the dot-dashed contours represent the method based in the mean value of the survival probabilities of the ALPs. Both methods where applied to the same HAWC dataset of M87. The CAST experiment's excluded region is shown in red for comparison. (For interpretation of the references to colour in this figure legend, the reader is referred to the web version of this article.)

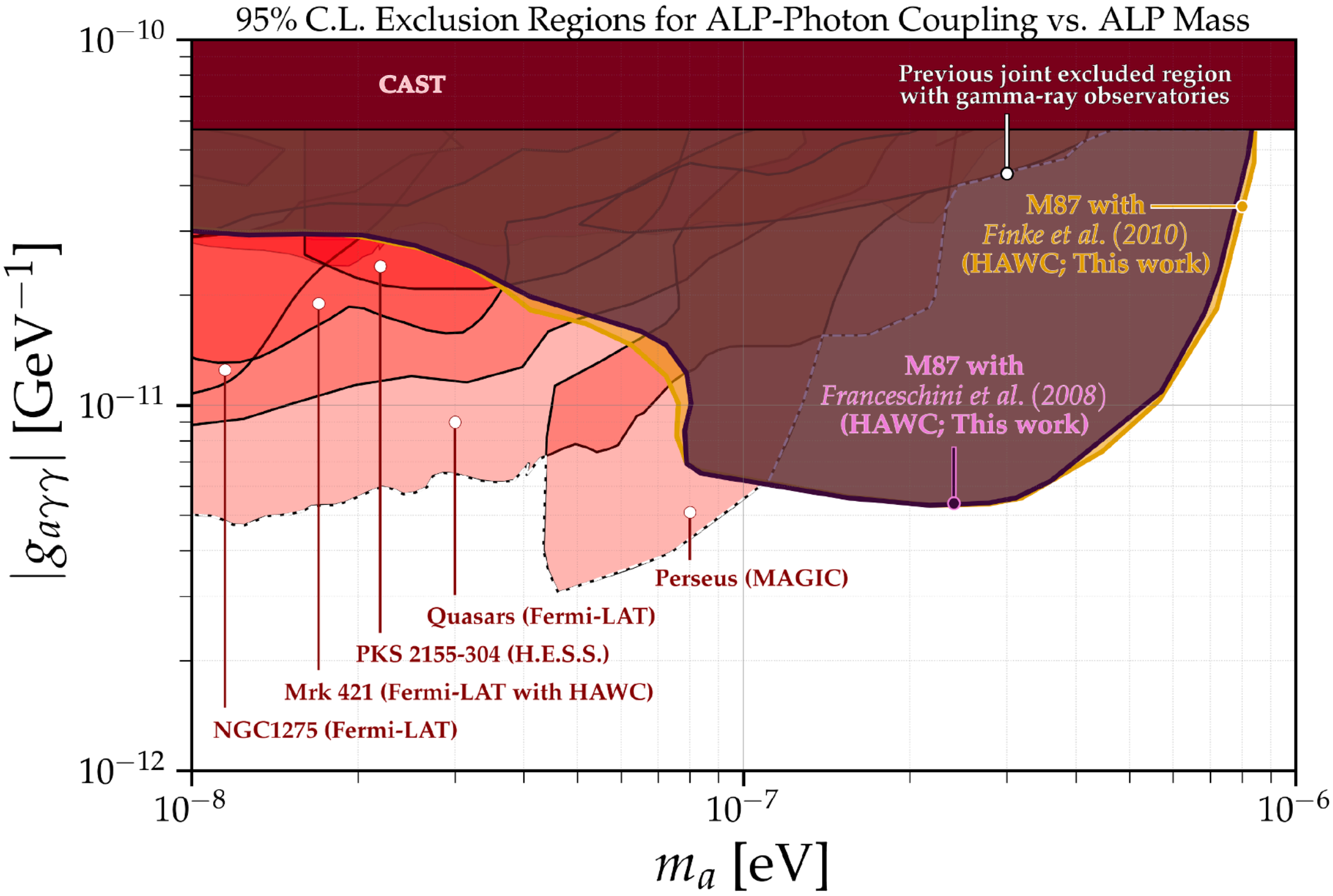

**Fig. 5.** Comparison of the 95% C.L. ALP exclusion regions derived in this work with previous constraints from VHE gamma-ray observations. The dark purple and gold regions represent the robust limits obtained for M87 using the Franceschini et al. (2008) [12] and Finke et al. (2010) [15] EBL models, respectively. Also shown are constraints from the CAST experiment (dark red) and various astrophysical sources including NGC 1275 (Fermi-LAT) [80], PKS 2155-304 (H.E.S.S.) [82], and the Perseus cluster (MAGIC) [83]. Our analysis provides complementary coverage, particularly in the mass regime above $10^{-7}$ eV compared to the other gamma-ray observatories' regions. (For interpretation of the references to colour in this figure legend, the reader is referred to the web version of this article.)

The novelty of this work lies in the unique combination of HAWC's continuous monitoring capabilities and its optimized sensitivity in the multi-TeV regime. Specifically, while IACTs offer excellent instantaneous sensitivity, HAWC's nearly 100% duty cycle allows for the accumulation of a deep, multi-year integrated exposure that is difficult to achieve with the pointed, clear-sky observation requirements of IACTs. Furthermore, HAWC's large effective area provides the necessary statistics to extend the M87 spectrum into the multi-TeV range, beyond the typical energy reach of space-based instruments like Fermi-LAT. Finally, while newer observatories such as LHAASO offer higher instantaneous sensitivity and efficiency at the highest energies, this HAWC dataset spans more than seven years of continuous observation, representing the largest integrated dataset of M87 at VHE to date. This makes HAWC an ideal and complementary tool for searching for the high-energy spectral "revival" predicted by ALP-photon mixing models. With the

The novelty of this work lies in the utilization of a 7.5-year continuous dataset, which currently represents the longest accumulated exposure of M87 in the multi-TeV regime. While IACTs offer higher angular and energy resolution, they are limited by low duty cycles and short observation windows. In contrast, the HAWC Observatory provides an unbiased, long-term spectral baseline that is less susceptible to transient flaring states, allowing for a more robust characterization of the source's typical emission profile. Furthermore, while the LHAASO observatory has demonstrated higher instantaneous sensitivity, the established 7.5-year HAWC archive offers a unique statistical advantage in defining the spectral shape at energies above 10 TeV without the sampling biases inherent to shorter-duration or pointed observations. This long-term monitoring, combined with our conservative intersection method for handling polarization uncertainties, provides robust and complementary constraints to the existing VHE literature.

Regarding the statistical framework, our approach diverges from methods that rely on the mean survival probability. Averaging over the ensemble of magnetic field realizations tends to smooth out specific spectral irregularities (wiggles) due to phase cancellations between different configurations. Since the actual physical Universe corresponds to a single realization rather than an average, such smoothing could mask a detectable signal. By instead defining exclusion regions based on the intersection of individual realizations, we preserve these specific spectral features, ensuring that the sensitivity to ALP signatures is not underestimated by artificial averaging.

## 7. Conclusions

This work demonstrates the value of long-term, wide-field observations in constraining fundamental physics beyond the Standard Model. The HAWC observatory capability to continuously monitor the sky, large FoV, and sensitivity to the highest energy gamma rays, make it well-suited for systematic searches of spectral anomalies that could reveal new physics. While the photon-ALP conversion mechanism remains undetected in our data, the methodology developed here, particularly the conservative treatment of polarization uncertainties through the intersection method, provides a robust framework applicable to future searches with both current and next-generation instruments.

The global effort to detect or constrain axion-like particles requires diverse approaches across multiple wavelengths, targets, and experimental techniques. Ground-based gamma-ray observatories occupy a crucial position in this subject, probing parameter space complementary to laboratory experiments, solar observations, and cosmological constraints. In this context, the constraints derived here provide a valu-able benchmark for upcoming next-generation facilities, such as the Cherenkov Telescope Array Observatory (CTAO) [84] and the Southern Wide-field Gamma-ray Observatory (SWGO) [85]. These new instruments will offer superior sensitivity at multi-TeV energies, enabling both deeper ALP searches and highly refined modeling of the astrophysical magnetic fields, thereby extending the legacy of constraints set by HAWC. As observational capabilities improve and theoretical models become more refined, even null results contribute essential information that shapes our understanding of viable dark matter candidates and guides future experimental priorities.

The path forward requires not only technological advances in detector sensitivity and energy resolution but also improved understanding of the astrophysical environments through which gamma-rays propagate. Progress in characterizing cosmic magnetic fields, reducing systematic uncertainties in EBL models, and developing more sophisticated analysis techniques will be as important as raw instrumental improvements. The continuation of M87 observations across multiple instruments and the identification of new, optimally suited targets will remain critical components of the ongoing search for these elusive particles that could fundamentally alter our understanding of dark matter and physics beyond the Standard Model.

## CRediT authorship contribution statement

**R. Alfaro:** Writing – review & editing, Validation, Supervision, Resources, Project administration, Investigation, Funding acquisition, Conceptualization; **C. Alvarez:** Writing – review & editing; **A. Andrés:** Writing – review & editing; **E. Anita-Rangel:** Writing – review & editing; **M. Araya :** Writing – review & editing; **J.C. Arteaga-Velázquez:** Writing – review & editing; **D. Avila Rojas :** Writing – review & editing, Writing – original draft, Visualization, Validation, Software, Methodology, Investigation, Formal analysis, Data curation; **H.A. Ayala Solares :** Writing – review & editing; **R. Babu :** Writing – review & editing; **P. Bangale :** Writing – review & editing; **E. Belmont-Moreno :** Writing – review & editing; **A. Bernal:** Writing – review & editing; **K.S. Caballero-Mora :** Writing – review & editing; **T. Capistrán :** Writing – review & editing; **A. Carramiñana :** Writing – review & editing; **F. Carreón:** Writing – review & editing; **S. Casanova :** Writing – review & editing; **U. Cotti :** Writing – review & editing; **J. Cotzomi :** Writing – review & editing; **P. Desiati:** Writing – review & editing; **E. De La Fuente :** Writing – review & editing; **C. De León :** Writing – review & editing; **N. Di Lalla:** Writing – review & editing; **R. Diaz Hernandez:** Writing – review & editing; **M.A. DuVernois :** Writing – review & editing; **J.C. Díaz-Vélez :** Writing – review & editing; **T. Ergin :** Writing – review & editing; **C. Espinoza :** Writing – review & editing; **N. Fraija :** Writing – review & editing; **S. Fraija:** Writing – review & editing; **J.A. García-González :** Writing – review & editing; **F. Garfias :** Writing – review & editing; **N. Ghosh:** Writing – review & editing; **A. Gonzalez Muñoz:** Writing – review & editing; **M.M. González :** Writing – review & editing, Validation, Supervision, Resources, Project administration, Investigation, Funding acquisition, Conceptualization; **J.A. González:** Writing – review & editing; **J.A. Goodman :** Writing – review & editing; **J. Gyeong:** Writing – review & editing; **J.P. Harding :** Writing – review & editing; **I. Herzog :** Writing – review & editing; **D. Huang :** Writing – review & editing; **F. Hueyotl-Zahuantitla :** Writing – review & editing; **A. Iriarte :** Writing – review & editing; **S. Kaufmann:** Writing – review & editing; **D. Kieda :** Writing – review & editing; **A. Lara :** Writing – review & editing; **W.H. Lee :** Writing – review & editing; **J. Lee :** Writing – review & editing; **H. León Vargas :** Writing – review & editing; **J.T. Linnemann :** Validation, Visualization, Writing – review & editing; **A.L. Longinotti :** Writing – review & editing; **G. Luis-Raya :** Writing – review & editing; **K. Malone :** Writing – review & editing; **O. Martinez :** Writing – review & editing; **J. Martínez-Castro :** Writing – review & editing; **H. Martínez-Huerta :** Writing – review & editing; **J.A. Matthews :** Writing – review & editing; **P. Miranda-Romagnoli :** Writing – review & editing; **P.E. Mirón-Enriquez :** Writing – review & editing; **J.A. Morales-Soto :** Writing – review & editing; **E. Moreno :** Writing – review & editing; **M. Mostafá :** Writing – review & editing; **M. Najafi:** Writing – review & editing; **A. Nayerhoda:** Writing – review & editing; **L. Nellen :** Writing – review & editing; **M.U. Nisa :** Writing – review & editing; **R. Noriega-Papaqui :** Writing – review & editing; **L. Olivera-Nieto :** Writing – review & editing; **N. Omodei :** Writing – review & editing; **M. Osorio-Archila:** Writing – review & editing; **E. Ponce:** Writing – review & editing; **Y. Pérez Araujo :** Writing – review & editing; **E.G. Pérez-Pérez :** Writing – review & editing; **A. Pratts:** Writing – review & editing, Writing – original draft, Visualization, Validation, Software, Methodology, Investigation, Formal analysis, Data curation, Conceptualization; **C.D. Rho :** Writing – review & editing; **A. Rodriguez Parra:** Writing – review & editing; **D. Rosa-González :** Writing – review & editing; **M. Roth:** Writing – review & editing; **D. Salazar-Gallegos:** Writing – review & editing; **A. Sandoval :** Writing – review & editing; **M. Schneider :** Writing – review & editing; **J. Serna-Franco :** Writing – review & editing, Writing – original draft, Visualization, Validation, Software, Methodology, Investigation, Formal analysis, Data curation, Conceptualization; **A.J. Smith :** Writing – review & editing; **Y. Son:** Writing – review & editing; **R.W. Springer :** Writing – review & editing; **O. Tibolla:** Writing – review & editing; **K. Tollefson :** Writing – review & editing; **I. Torres :** Writing – review & editing; **R. Torres-Escobedo :** Writing – review & editing; **E. Varela:** Writing – review & editing; **L. Villaseñor :** Writing – review & editing; **X. Wang :** Writing – review & editing; **Z. Wang:** Writing – review & editing; **I.J. Watson :** Writing – review & editing; **H. Wu:** Writing – review & editing; **S. Yu:** Writing – review & editing; **X. Zhang:** Writing – review & editing; **H. Zhou :** Writing – review & editing.

## Data availability

The authors do not have permission to share data.

## Declaration of competing interest

The authors declare the following financial interests/personal relationships which may be considered as potential competing interests:

M.M. Gonzalez, J. Serna-Franco, R. Alfaro, D. Avila Rojas, A. Pratts reports financial support was provided by National Autonomous University of Mexico Directorate General of Academic Staff Affairs. If there are other authors, they declare that they have no known competing financial interests or personal relationships that could have appeared to influence the work reported in this paper.

## Acknowledgments

We acknowledge the support from: the US National Science Foundation (NSF); the US Department of Energy Office of High-Energy Physics; the Laboratory Directed Research and Development (LDRD) program of Los Alamos National Laboratory; Consejo Nacional de Ciencia y Tecnología (CONACyT), México, grants LNC-2023-117, 271051, 232656, 260378, 179588, 254964, 258865, 243290, 132197, A1-S-46288, A1-S-22784, CF-2023-I-645, cátedras 873, 1563, 341, 323, Red HAWC, México; DGAPA-UNAM grants IG101323, IN111716-3, IN111419, IA102019, IN106521, IN114924, IN110521, IN102223, IG100726; Laboratorios Nacionales SECIHTI LNC-2023-117; VIEP-BUAP; PIFI 2012, 2013, PROFOCIE 2014, 2015; the University of Wisconsin Alumni Research Foundation; the Institute of Geophysics, Planetary Physics, and Signatures at Los Alamos National Laboratory; Polish Science Centre grant, 2024/53/B/ST9/02671; Coordinación de la Investigación Científica de la Universidad Michoacana; Royal Society - Newton Advanced Fellowship 180385; Gobierno de España and European Union-NextGenerationEU, grant CNS2023- 144099; The Program Management Unit for Human Resources & Institutional Development, Research and Innovation, NXPO (grant number B16F630069); Coordinación General Académica e Innovación (CGAI-UdeG), PRODEP-SEP UDG-CA-499; Institute of Cosmic Ray Research (ICRR), University of Tokyo. H.F. ac-knowledges support by NASA under award number 80GSFC21M0002. C.R. acknowledges support from National Research Foundation of Korea (RS-2023-00280210). We also acknowledge the significant contributions over many years of Stefan Westerhoff, Gaurang Yodh and Arnulfo Zepeda Domínguez, all deceased members of the HAWC collaboration. Thanks to Scott Delay, Luciano Díaz and Eduardo Murrieta for technical support.

## References

[1] P.J.E. Peebles, Dark matter and the origin of galaxies and globular star clusters, Astrophys. J. 277 (1984) 470–477. Provided by the SAO/NASA Astrophysics Data System, https://ui.adsabs.harvard.edu/abs/1984ApJ\protect\LY1\textellipsis277.470P. https://doi.org/10.1086/161714
[2] P.J. Peebles, B. Ratra, The cosmological constant and dark energy, Rev. Mod. Phys. 75 (2) (2003) 559–606. Provided by the SAO/NASA Astrophysics Data System, arXiv:astro-ph/0207347. https://doi.org/10.1103/RevModPhys.75.559
[3] R.D. Peccei, H.R. Quinn, CP Conservation in the presence of pseudoparticles, Phys. Rev. Lett. 38 (25) (1977) 1440–1443. Provided by the SAO/NASA Astrophysics Data System, https://doi.org/10.1103/PhysRevLett.38.1440
[4] R.D. Peccei, H.R. Quinn, Constraints imposed by CP conservation in the presence of pseudoparticles, Phys. Rev. D 16 (6) (1977) 1791–1797. Provided by the SAO/NASA Astrophysics Data System, https://doi.org/10.1103/PhysRevD.16.1791
[5] R.D. Peccei, The strong CP problem and axions, Provided by the SAO/NASA Astrophysics Data System, 741 (2008) 3. https://ui.adsabs.harvard.edu/abs/2008LNP\protect\LY1\textellipsis741\protect\LY1\textellipsis.3P. https://doi.org/10.1007/978-3-540-73518-2_1
[6] G. Raffelt, et al., Mixing of the photon with low-mass particles, Phys. Rev. D 37 (5) (1988) 1237–1249. Provided by the SAO/NASA Astrophysics Data System, https://doi.org/10.1103/PhysRevD.37.1237
[7] L. Hui, J.P. Ostriker, S. Tremaine, E. Witten, Ultralight scalars as cosmological dark matter, Phys. Rev. D 95 (4) (2017) 043541. Provided by the SAO/NASA Astrophysics Data System, arXiv:1610.08297. https://doi.org/10.1103/PhysRevD.95.043541
[8] D.J.E. Marsh, Axion Cosmology, arXiv e-prints (2015) Provided by the SAO/NASA Astrophysics Data System, arXiv:1510.07633. https://doi.org/10.48550/arXiv.1510.07633
[9] D. Cadamuro, J. Redondo, Cosmological bounds on pseudo Nambu-Goldstone bosons, J. Cosmol. Astropart. Phys. 2012 (2) (2012) 032. Provided by the SAO/NASA Astrophysics Data System, arXiv:1110.2895. https://doi.org/10.1088/1475-7516/2012/02/032
[10] J. Jaeckel, M. Spannowsky, Probing MeV to 90 GeV axion-like particles with LEP and LHC, Phys. Lett. B 753 (2016) 482–487. Provided by the SAO/NASA Astrophysics Data System, arXiv:1509.00476. https://doi.org/10.1016/j.physletb.2015.12.037
[11] A. Mirizzi, D. Montanino, Stochastic conversions of TeV photons into axion-like particles in extragalactic magnetic fields, J. Cosmol. Astropart. Phys. 2009 (12) (2009) 004. Provided by the SAO/NASA Astrophysics Data System, arXiv:0911.0015. https://doi.org/10.1088/1475-7516/2009/12/004
[12] A. Franceschini, et al., Extragalactic optical-infrared background radiation, its time evolution and the cosmic photon-photon opacity, Astron. Astrophys. 487 (3) (2008) 837–852. Provided by the SAO/NASA Astrophysics Data System, arXiv:0805.1841. https://doi.org/10.1051/0004-6361:200809691
[13] A. Franceschini, G. Rodighiero, The extragalactic background light revisited and the cosmic photon-photon opacity, Astron. Astrophys. 603 (2017) A34. Provided by the SAO/NASA Astrophysics Data System, arXiv:1705.10256. https://doi.org/10.1051/0004-6361/201629684
[14] A. Domínguez, et al., Extragalactic background light inferred from AEGIS galaxy-SED-type fractions, Mon. Not. R. Astron. Soc. 410 (4) (2011) 2556–2578. Provided by the SAO/NASA Astrophysics Data System, arXiv:1007.1459. https://doi.org/10.1111/j.1365-2966.2010.17631.x
[15] J.D. Finke, et al., Modeling the Extragalactic Background Light from Stars and Dust, Astrophys. J. 712 (1) (2010) 238–249. Provided by the SAO/NASA Astrophysics Data System, arXiv:0905.1115. https://doi.org/10.1088/0004-637X/712/1/238
[16] M. Meyer, et al., On detecting oscillations of gamma rays into axion-like particles in turbulent and coherent magnetic fields, J. Cosmol. Astropart. Phys. 2014 (9) (2014) 003–003. Provided by the SAO/NASA Astrophysics Data System, arXiv:1406.5972. https://doi.org/10.1088/1475-7516/2014/09/003
[17] P. Brun, D. Wouters, Constraints on Axion-like Particles with H.E.S.S. from Observations of PKS 2155-304, in: International Cosmic Ray Conference, 33 of International Cosmic Ray Conference, 2013, p. 2733. Provided by the SAO/NASA Astrophysics Data System, arXiv:1307.6068. https://doi.org/10.48550/arXiv.1307.6068
[18] M. Ajello, et al., Search for Spectral Irregularities due to Photon-Axionlike-Particle Oscillations with the Fermi Large Area Telescope, Phys. Rev. Lett. 116 (16) (2016) 161101. Provided by the SAO/NASA Astrophysics Data System, arXiv:1603.06978. https://doi.org/10.1103/PhysRevLett.116.161101
[19] A. Abramowski, et al., Constraints on axionlike particles with H.E.S.S. from the irregularity of the PKS 2155-304 energy spectrum, Phys. Rev. D 88 (10) (2013) 102003. Provided by the SAO/NASA Astrophysics Data System, arXiv:1311.3148. https://doi.org/10.1103/PhysRevD.88.102003
[20] S. Mei, et al., The ACS Virgo Cluster Survey. XIII. SBF Distance Catalog and the Three-dimensional Structure of the Virgo Cluster, Astrophys. J. 655 (1) (2007) 144–162. Provided by the SAO/NASA Astrophysics Data System, arXiv:astro-ph/0702510. https://doi.org/10.1086/509598
[21] R. Alfaro, et al., Longtime Monitoring of TeV Radio Galaxies with HAWC, arXiv e-prints (2025) Provided by the SAO/NASA Astrophysics Data System, arXiv:2506.16031. https://doi.org/10.48550/arXiv.2506.16031
[22] Z. Cao, et al., Detection of Very High-energy Gamma-Ray Emission from the Radio Galaxy M87 with LHAASO, Astrophys. J. Lett. 975 (2) (2024) L44. Provided by the SAO/NASA Astrophysics Data System, arXiv:2410.15353. https://doi.org/10.3847/2041-8213/ad8921
[23] D.E. Harris, et al., Variability Timescales in the M87 Jet: Signatures of E $^2$ Losses, Discovery of a Quasi Period in HST-1, and the Site of TeV Flaring, Astrophys. J. 699 (1) (2009) 305–314. Provided by the SAO/NASA Astrophysics Data System, arXiv:0904.3925. https://doi.org/10.1088/0004-637X/699/1/305
[24] O. Urban, et al., A uniform metallicity in the outskirts of massive, nearby galaxy clusters, Mon. Not. R. Astron. Soc. 470 (4) (2017) 4583–4599. Provided by the SAO/NASA Astrophysics Data System, arXiv:1706.01567. https://doi.org/10.1093/mnras/stx1542
[25] O. Urban, et al., X-ray spectroscopy of the Virgo Cluster out to the virial radius, Mon. Not. R. Astron. Soc. 414 (3) (2011) 2101–2111. Provided by the SAO/NASA Astrophysics Data System, arXiv:1102.2430. https://doi.org/10.1111/j.1365-2966.2011.18526.x
[26] R. Cecil, et al., Probing gamma-Ray propagation at very-High energies with h.e.s.s. observations of M87, in: 38th International Cosmic Ray Conference, 2024, p. 908. Provided by the SAO/NASA Astrophysics Data System, https://ui.adsabs.harvard.edu/abs/2024icrc.confE.908C.
[27] C.L. Carilli, G.B. Taylor, Cluster Magnetic Fields, Annu. Rev. Astron. Astrophys. 40 (2002) 319–348. Provided by the SAO/NASA Astrophysics Data System, arXiv:astro-ph/0110655. https://doi.org/10.1146/annurev.astro.40.060401.093852
[28] F. Govoni, L. Feretti, Magnetic Fields in Clusters of Galaxies, Int. J. Mod. Phys. D 13 (8) (2004) 1549–1594. Provided by the SAO/NASA Astrophysics Data System, arXiv:astro-ph/0410182. https://doi.org/10.1142/S0218271804005080
[29] E.H.T. Collaboration, First M87 Event Horizon Telescope Results. I. The Shadow of the Supermassive Black Hole, Astrophys. J. Lett. 875 (1) (2019) L1. Provided by the SAO/NASA Astrophysics Data System, arXiv:1906.11238. https://doi.org/10.3847/2041-8213/ab0ec7
[30] E.H.T. Collaboration, First M87 Event Horizon Telescope Results. II. Array and Instrumentation, Astrophys. J. Lett. 875 (1) (2019) L2. Provided by the SAO/NASA Astrophysics Data System, arXiv:1906.11239. https://doi.org/10.3847/2041-8213/ab0c96
[31] E.H.T. Collaboration, First M87 Event Horizon Telescope Results. III. Data Processing and Calibration, Astrophys. J. Lett. 875 (1) (2019) L3. Provided by the SAO/NASA Astrophysics Data System, arXiv:1906.11240. https://doi.org/10.3847/2041-8213/ab0c57
[32] B.C. Whitmore, et al., Hubble space telescope observations of globular clusters in M87 and an estimate of h 0, Astrophys. J. Lett. 454 (1995) L73. Provided by the SAO/NASA Astrophysics Data System, https://doi.org/10.1086/309788
[33] A.S. Wilson, Y. Yang, Chandra X-Ray Imaging and Spectroscopy of the M87 Jet and Nucleus, Astrophys. J. 568 (1) (2002) 133–140. Provided by the SAO/NASA Astrophysics Data System, arXiv:astro-ph/0112097. https://doi.org/10.1086/338887
[34] W. Forman, C. Jones, E. Churazov, M. Markevitch, P. Nulsen, A. Vikhlinin, M. Begelman, H. Bøhringer, J. Eilek, S. Heinz, R. Kraft, F. Owen, M. Pahre, Filaments, Bubbles, and Weak Shocks in the Gaseous Atmosphere of M87, Astrophys. J. 665 (2) (2007) 1057–1066. Provided by the SAO/NASA Astrophysics Data System, arXiv:astro-ph/0604583. https://doi.org/10.1086/519480
[35] M. Boughelilba, et al., Lepto-hadronic Jet-disk Model for the Multiwavelength SED of M87, Astrophys. J. 938 (1) (2022) 79. Provided by the SAO/NASA Astrophysics Data System, arXiv:2208.14756. https://doi.org/10.3847/1538-4357/ac8e64
[36] A. Abramowski, et al., The 2010 Very High Energy $\gamma$-Ray Flare and 10 Years of Multi-wavelength Observations of M 87, Astrophys. J. 746 (2) (2012) 151. Provided by the SAO/NASA Astrophysics Data System, arXiv:1111.5341. https://doi.org/10.1088/0004-637X/746/2/151
[37] V.A. Acciari, et al., Radio Imaging of the Very-High-Energy $\gamma$-Ray Emission Region in the Central Engine of a Radio Galaxy, Science 325 (5939) (2009) 444. Provided by the SAO/NASA Astrophysics Data System, arXiv:0908.0511. https://doi.org/10.1126/science.1175406
[38] F. Aharonian, et al., Is the giant radio galaxy M 87 a TeV gamma-ray emitter?, Astron. Astrophys. 403 (2003) L1–L5. Provided by the SAO/NASA Astrophysics Data System, arXiv:astro-ph/0302155. https://doi.org/10.1051/0004-6361:20030372
[39] F. Aharonian, et al., Fast Variability of Tera-Electron Volt $\gamma$ Rays from the Radio Galaxy M87, Science 314 (5804) (2006) 1424–1427. Provided by the SAO/NASA Astrophysics Data System, arXiv:astro-ph/0612016. https://doi.org/10.1126/science.1134408
[40] C. Arcaro, et al., Probing the morphology of the low state gamma-ray emission of M87 with h.e.s.s, in: 38th International Cosmic Ray Conference, 2024, p. 696. Provided by the SAO/NASA Astrophysics Data System, https://ui.adsabs.harvard.edu/abs/2024icrc.confE.696A.
[41] M. Collaboration, Monitoring of the radio galaxy M 87 during a low-emission state from 2012 to 2015 with MAGIC, Mon. Not. R. Astron. Soc. 492 (4) (2020) 5354–5365. Provided by the SAO/NASA Astrophysics Data System, arXiv:2001.01643. https://doi.org/10.1093/mnras/staa014
[42] M. Molero Gonzalez, et al., Long-term monitoring of the radio-galaxy M87 in gamma-rays: joint analysis of MAGIC, VERITAS and fermi-LAT data, in: 38th International Cosmic Ray Conference, 2024, p. 572. Provided by the SAO/NASA Astrophysics Data System, https://ui.adsabs.harvard.edu/abs/2024icrc.confE.572M.

[43] J. Aleksić, et al., MAGIC observations of the giant radio galaxy M 87 in a low-emission state between 2005 and 2007, Astron. Astrophys. 544 (2012) A96. Provided by the SAO/NASA Astrophysics Data System, arXiv:1207.2147. https://doi.org/10.1051/0004-6361/201117827

[44] M. Beilicke, et al., VERITAS observations of M87 in 2011/2012, in: F.A. Aharonian, W. Hofmann, F.M. Rieger (Eds.), High Energy Gamma-Ray Astronomy: 5th International Meeting on High Energy Gamma-Ray Astronomy, 1505 of American Institute of Physics Conference Series, AIP, 2012, pp. 586–589. Provided by the SAO/NASA Astrophysics Data System, arXiv:1210.7830. https://doi.org/10.1063/1.4772328

[45] A.U. Abeysekara, et al., Observation of the Crab Nebula with the HAWC Gamma-Ray Observatory, Astrophys. J. 843 (1) (2017) 39. Provided by the SAO/NASA Astrophysics Data System, arXiv:1701.01778. https://doi.org/10.3847/1538-4357/aa7555

[46] A.U. Abeysekara, et al., The High-Altitude Water Cherenkov (HAWC) observatory in México: The primary detector, Nucl. Instrum. Methods Phys. Res. A 1052 (2023) 168253. Provided by the SAO/NASA Astrophysics Data System, arXiv:2304.00730. https://doi.org/10.1016/j.nima.2023.168253

[47] R. Alfaro, et al., Study of the Very High Energy Emission of M87 through its Broadband Spectral Energy Distribution, Astrophys. J. 934 (2) (2022) 158. Provided by the SAO/NASA Astrophysics Data System, arXiv:2112.09179. https://doi.org/10.3847/1538-4357/ac7b78

[48] A. Albert, et al., Performance of the HAWC Observatory and TeV Gamma-Ray Measurements of the Crab Nebula with Improved Extensive Air Shower Reconstruction Algorithms, Astrophys. J. 972 (2) (2024) 144. Provided by the SAO/NASA Astrophysics Data System, arXiv:2405.06050. https://doi.org/10.3847/1538-4357/ad5f2d

[49] A. Albert, et al., Search for decaying dark matter in the Virgo cluster of galaxies with HAWC, Phys. Rev. D 109 (4) (2024) 043034. Provided by the SAO/NASA Astrophysics Data System, arXiv:2309.03973. https://doi.org/10.1103/PhysRevD.109.043034

[50] S.M. Faber, J.S. Gallagher, Masses and mass-to-light ratios of galaxies, Annu. Rev. Astron. Astrophys. 17 (1979) 135–187. Provided by the SAO/NASA Astrophysics Data System, https://doi.org/10.1146/annurev.aa.17.090179.001031

[51] K. Gebhardt, J. Thomas, The Black Hole Mass, Stellar Mass-to-Light Ratio, and Dark Halo in M87, Astrophys. J. 700 (2) (2009) 1690–1701. Provided by the SAO/NASA Astrophysics Data System, arXiv:0906.1492. https://doi.org/10.1088/0004-637X/700/2/1690

[52] Z. Wang, A.E. Broderick, Constraining the Existence of Axion Clouds in M87* with Closure Trace Analyses, Astrophys. J. 962 (2) (2024) 121. Provided by the SAO/NASA Astrophysics Data System, arXiv:2311.01565. https://doi.org/10.3847/1538-4357/ad13f4

[53] A. Kar, et al., Constraining eV-scale axion-like particle dark matter: insights from the M87 Galaxy, arXiv e-prints (2025) Provided by the SAO/NASA Astrophysics Data System, arXiv:2501.01860. https://doi.org/10.48550/arXiv.2501.01860

[54] A. Pratts, et al., Axion-like Particles and their Possible Impact on the Very High-Energy Spectrum of M87 Observed by LHAASO, arXiv e-prints (2025) Provided by the SAO/NASA Astrophysics Data System, arXiv:2507.05637. https://doi.org/10.48550/arXiv.2507.05637

[55] V. Joshi, A. Jardin-Blicq, et al., HAWC High Energy Upgrade with a Sparse Outrigger Array, in: 35th International Cosmic Ray Conference (ICRC2017), 301 of International Cosmic Ray Conference, 2017, p. 806. Provided by the SAO/NASA Astrophysics Data System, arXiv:1708.04032. https://doi.org/10.22323/1.301.0806

[56] A.U. Abeysekara, et al., Measurement of the Crab Nebula Spectrum Past 100 TeV with HAWC, Astrophys. J. 881 (2) (2019) 134. Provided by the SAO/NASA Astrophysics Data System, arXiv:1905.12518. https://doi.org/10.3847/1538-4357/ab2f7d

[57] D. Avila Rojas, T. Capistrán, M.M. Gonzalez, R.J. Alfaro, J. Serna-Franco, M. Osorio, HAWC Collaboration, Studies on the VHE spectrum of the radio galaxy M87 observed by HAWC, in: 39th International Cosmic Ray Conference, 2026, p. 789. Provided by the SAO/NASA Astrophysics Data System,

[58] A. Donath, et al., Gammapy: A Python package for gamma-ray astronomy, Astron. Astrophys. 678 (2023) A157. Provided by the SAO/NASA Astrophysics Data System, arXiv:2308.13584. https://doi.org/10.1051/0004-6361/202346488

[59] A. Saldana-Lopez, et al., An observational determination of the evolving extragalactic background light from the multiwavelength HST/CANDELS survey in the Fermi and CTA era, Mon. Not. R. Astron. Soc. 507 (4) (2021) 5144–5160. Provided by the SAO/NASA Astrophysics Data System, arXiv:2012.03035. https://doi.org/10.1093/mnras/stab2393

[60] G. Vianello, et al., The Multi-Mission Maximum Likelihood framework (3ML), arXiv e-prints (2015) Provided by the SAO/NASA Astrophysics Data System, arXiv:1507.08343. https://doi.org/10.48550/arXiv.1507.08343

[61] A. Albert, et al., Validation of standardized data formats and tools for ground-level particle-based gamma-ray observatories, Astron. Astrophys. 667 (2022) A36. Provided by the SAO/NASA Astrophysics Data System, arXiv:2203.05937. https://doi.org/10.1051/0004-6361/202243527

[62] A. de Angelis, et al., Relevance of axionlike particles for very-high-energy astrophysics, Phys. Rev. D 84 (10) (2011) 105030. Provided by the SAO/NASA Astrophysics Data System, arXiv:1106.1132. https://doi.org/10.1103/PhysRevD.84.105030

[63] X. Bi, et al., Axion and dark photon limits from Crab Nebula high-energy gamma rays, Phys. Rev. D 103 (4) (2021) 043018. Provided by the SAO/NASA Astrophysics Data System, arXiv:2002.01796. https://doi.org/10.1103/PhysRevD.103.043018

[64] D. Wouters, P. Brun, Irregularity in gamma ray source spectra as a signature of axion-like particles, Phys. Rev. D 86 (4) (2012) 043005. Provided by the SAO/NASA Astrophysics Data System, arXiv:1205.6428. https://doi.org/10.1103/PhysRevD.86.043005

[65] R. Jansson, G.R. Farrar, A New Model of the Galactic Magnetic Field, Astrophys. J. 757 (1) (2012) 14. Provided by the SAO/NASA Astrophysics Data System, arXiv:1204.3662. https://doi.org/10.1088/0004-637X/757/1/14

[66] R.C. Walker, et al., The Structure and Dynamics of the Subparsec Jet in M87 Based on 50 VLBA Observations over 17 Years at 43 GHz, Astrophys. J. 855 (2) (2018) 128. Provided by the SAO/NASA Astrophysics Data System, arXiv:1802.06166. https://doi.org/10.3847/1538-4357/aaafcc

[67] M. Meyer, et al., gammaALPs: An open-source python package for computing photon-axion-like-particle oscillations in astrophysical environments, in: 37th International Cosmic Ray Conference, 2022, p. 557. Provided by the SAO/NASA Astrophysics Data System, arXiv:2108.02061. https://doi.org/10.22323/1.395.0557

[68] SIMBAD Astronomical Database, M 87, 2025, (2025). Access: Feb 9, 2025, https://simbad.cds.unistra.fr/simbad/simbasicIdent = M + 87&submit = SIMBAD + search.

[69] A.S. Nikonov, et al., Properties of the jet in M87 revealed by its helical structure imaged with the VLBA at 8 and 15 GHz, Mon. Not. R. Astron. Soc. 526 (4) (2023) 5949–5963. Provided by the SAO/NASA Astrophysics Data System, arXiv:2307.11660. https://doi.org/10.1093/mnras/stad3061

[70] A. Pasetto, et al., Reading M87's DNA: A Double Helix Revealing a Large-scale Helical Magnetic Field, Astrophys. J. Lett. 923 (1) (2021) L5. Provided by the SAO/NASA Astrophysics Data System, arXiv:2112.06971. https://doi.org/10.3847/2041-8213/ac3a88

[71] O.G. Kashibadze, et al., Structure and kinematics of the Virgo cluster of galaxies, Astron. Astrophys. 635 (2020) A135. Provided by the SAO/NASA Astrophysics Data System, arXiv:2002.12820. https://doi.org/10.1051/0004-6361/201936172

[72] W.J. Potter, G. Cotter, New constraints on the structure and dynamics of black hole jets, Mon. Not. R. Astron. Soc. 453 (4) (2015) 4070–4088. Provided by the SAO/NASA Astrophysics Data System, arXiv:1508.00567. https://doi.org/10.1093/mnras/stv1657

[73] M.M. González, et al., GRB 221009A: A Light Dark Matter Burst or an Extremely Bright Inverse Compton Component?, Astrophys. J. 944 (2) (2023) 178. Provided by the SAO/NASA Astrophysics Data System, arXiv:2210.15857. https://doi.org/10.3847/1538-4357/acb700

[74] R. Cecil, M. Meyer, Modeling the magnetic field of the virgo cluster for axion like particle search in gamma ray energies, in: 7th Heidelberg International Symposium on High-Energy Gamma-Ray Astronomy, 2024, p. 182. Provided by the SAO/NASA Astrophysics Data System, https://ui.adsabs.harvard.edu/abs/2024hegr.confE.182C.

[75] H. Abe, et al., Constraints on axion-like particles with the Perseus Galaxy Cluster with MAGIC, Phys. Dark Universe 44 (2024) 101425. Provided by the SAO/NASA Astrophysics Data System, arXiv:2401.07798. https://doi.org/10.1016/j.dark.2024.101425

[76] S.S. Wilks, The large-sample distribution of the likelihood ratio for testing composite hypotheses, Ann. Math. Stat. 9 (1) (1938) 60–62. https://doi.org/10.1214/aoms/1177732360

[77] W.A. Rolke, et al., Limits and confidence intervals in the presence of nuisance parameters, Nucl. Instrum. Methods Phys. Res. A 551 (2–3) (2005) 493–503. Provided by the SAO/NASA Astrophysics Data System, arXiv:physics/0403059. https://doi.org/10.1016/j.nima.2005.05.068

[78] M. Ajello, et al., Search for Spectral Irregularities due to Photon-Axionlike-Particle Oscillations with the Fermi Large Area Telescope, Phys. Rev. Lett. 116 (16) (2016) 161101. Provided by the SAO/NASA Astrophysics Data System, arXiv:1603.06978. https://doi.org/10.1103/PhysRevLett.116.161101

[79] J. Serna-Franco, et al., Searching for ALPs signatures in the M87 spectrum with HAWC, in: 39th International Cosmic Ray Conference (ICRC2025), International Cosmic Ray Conference, 2025, p. 497. https://indico.cern.ch/event/1258933/contributions/6475958/.

[80] M. Ajello, et al., Search for Spectral Irregularities due to Photon-Axionlike-Particle Oscillations with the Fermi Large Area Telescope, Phys. Rev. Lett. 116 (16) (2016) 161101. Provided by the SAO/NASA Astrophysics Data System, arXiv:1603.06978. https://doi.org/10.1103/PhysRevLett.116.161101

[81] J. Davies, et al., Constraints on axionlike particles from a combined analysis of three flaring F e r m i flat-spectrum radio quasars, Phys. Rev. D 107 (8) (2023) 083027. Provided by the SAO/NASA Astrophysics Data System, arXiv:2211.03414. https://doi.org/10.1103/PhysRevD.107.083027

[82] D. Wouters, P. Brun, Constraints on axion-like particles from gamma-ray astronomy with H.E.S.S, arXiv e-prints (2013) Provided by the SAO/NASA Astrophysics Data System, arXiv:1304.0700. https://doi.org/10.48550/arXiv.1304.0700

[83] H. Abe, et al., Constraints on axion-like particles with the Perseus Galaxy Cluster with MAGIC, Phys. Dark Universe 44 (2024) 101425. Provided by the SAO/NASA Astrophysics Data System, arXiv:2401.07798. https://doi.org/10.1016/j.dark.2024.101425

[84] Cherenkov Telescope Array Consortium, et al., Science with the Cherenkov Telescope Array, World Scientific, 2019. Provided by the SAO/NASA Astrophysics Data System, https://doi.org/10.1142/10986

[85] SWGO Collaboration, et al., Science Prospects for the Southern Wide-field Gamma-ray Observatory: SWGO, arXiv e-prints (2025) Provided by the SAO/NASA Astrophysics Data System, arXiv:2506.01786. https://doi.org/10.48550/arXiv.2506.01786